\documentclass[]{aa}
\usepackage{graphicx}

\def\gta{ \lower .75ex \hbox{$\sim$} \llap{\raise .27ex \hbox{$>$}} }
\def\lta{ \lower .75ex\hbox{$\sim$} \llap{\raise .27ex \hbox{$<$}} }
\newcommand{\aap}{A\&A}
\newcommand{\apj}{ApJ}
\newcommand{\apjs}{ApJS}
\newcommand{\apjl}{ApJ Letter}
\newcommand{\mnras}{MNRAS}
\newcommand{\apss}{A\&ASS}

\begin{document}

\newcommand{\dt}[1]{{\frac{\partial #1}{\partial t}}}
\newcommand{\tin}[1]{{\mbox{\tiny{#1}}}}
\newcommand{\dpp}[1]{{\frac{\partial #1}{\partial p}}}
\newcommand{\dg}[1]{{\frac{\partial #1}{\partial \gamma}}}
\newcommand{\dppr}[1]{{\frac{\partial #1}{\partial p'}}}
\newcommand{\dx}[1]{{\frac{\partial #1}{\partial x}}}
\newcommand{\dtx}[1]{{\frac{\partial \widetilde{#1}}{\partial
      \widetilde{x}}}}
\newcommand{\dttx}[0]{{\frac{\partial}{\partial
      \widetilde{x}}}}
\newcommand{\wt}[1]{{\widetilde{#1}}}
\newcommand{\A}[0]{{\bf A}(p) }
\newcommand{\B}[0]{{\bf B}_{\pm}(p) }
\newcommand{\C}[0]{{\bf C}_{\pm}(p) }
\newcommand{\Rgg}[0]{{R}_{\gamma\gamma}}
\newcommand{\wtno}[0]{{\wt{n}}_{0}}
\newcommand{\wtRgg}[0]{{\wt{R}}_{\gamma\gamma}}
\newcommand{\tgg}[0]{{\tau}_{\gamma\gamma}}

\title{Pair creation at shocks:\\
 application to the high energy emission of compact objects}
\author{P.O. Petrucci \inst{1} \and G. Henri \inst{2} \and G. Pelletier \inst{2}}
\institute{Osservatorio Astronomico di Brera, via Brera 28, 20121 Milano,
Italy\\
\and 
Laboratoire d'Astrophysique, Observatoire de Grenoble,B.P 53X,
F38041 Grenoble Cedex, France}
\offprints{P.O. Petrucci}
\mail{petrucci@brera.mi.astro.it, petrucci@obs.ujf-grenoble.fr}
\date{Received ??/ accepted ??}
\abstract{
  We investigate the effect of pair creation on a shock structure.
  Particles, accelerated in the shock via the first order Fermi process,
  are supposed to cool by inverse Compton process on external soft
  photons, resulting in a cut-off power law shape of the particle
  distribution function. The high energy photons produced are thus able
  to create pairs, through photon-photon annihilation. The increase of
  the pair pressure may then be sufficient to modify the shock profile.
  We show that there is even a limit of the pair pressure (of the order
  of 20\% of the ram pressure of the upstream flow) above which the shock
  cannot exist any longer.\\ Conversely, significant changes of the flow
  velocity profile will also modify the spectral index and the high
  energy cut--off of the particle distribution function. Hence the number
  of particles able to trigger the pair creation process will change,
  modifying the pair creation rate accordingly. Taking into account these
  different processes, we self-consistently derive the flow velocity
  profile and the particle distribution function. We show that, in some
  region of the parameter space, the system can converge towards
  stationary states where pair creation and hydrodynamical effects
  balance.\\ We discuss the application of this model to explain the high
  energy emission observed in compact objects.  We show that hard X-ray
  spectra ($\alpha_{X}< 1.$) are only obtained for small pair pressure
  and we don't expect any strong annihilation line in this case. We
  suggest also a possible variability mechanism if the soft photon
  compactness depends itself on the pair density of the hot plasma, such
  as expected in reillumination models.
\keywords{Acceleration of particles -- Radiation mechanisms: non-thermal
  -- Shock waves -- X-rays: general}}
\titlerunning{Pair creation at shocks}
\authorrunning{P.O. Petrucci et al.}
\maketitle

\section{Introduction}
It is well-known that particles can be accelerated up to very high
energies by crossing a magnetized shock front many times and thus
experiencing the so-called first order Fermi process (Axford et al.,
\cite{axf76}; Krimsky \cite{kry77}; Bell \cite{bel78}; Blandford \&
Ostriker \cite{bla78}). The simple linear theory may be however
insufficient to describe a realistic situation because the accelerated
particles may induce a strong non-linear back reaction on the shock
itself. For example they may produce magnetohydrodynamic (Jokipii
\cite{jok76}; Lacombe \cite{lac77}) or electrostatic (Nishikawa et al.
\cite{nis94}) instabilities, and also strongly alter the hydrodynamics of
the shock through their own pressure (Drury et al. \cite{dru82};
Schneider \& Kirk \cite{sch87}).\\

In the case of astrophysical objects like the Active Galactic Nuclei
(herafter AGNs), where shocks are particularly attractive processes to
produce in situ high energy particles, the presence of important
radiation fields requires also to take into account particles-photons
interactions in the shock region. A large number of works have already
studied particles acceleration at shocks including radiative cooling
processes (Webb et al. \cite{web84}; Schlickeiser \cite{sch84}; Bregman
\cite{bre85}; Heavens \& Meisenheimer \cite{hea87}; Biermann \&
Strittmater \cite{bie87}; Protheroe \& Stanev \cite{pro99}; Drury et
al. \cite{dru99}) or the feedback of the accelerated particle pressures
on the shock structure (Blandford \cite{bla80}; Drury \& Volk
\cite{dru81}; Heavens \cite{hea83}).\\

The aim of this paper is to investigate also the effect of pair creation
on the shock structure. Indeed particles accelerated by the shock can be
sufficiently energetic to boost, via Inverse Compton (hereafter IC)
process for example, surrounding external soft photons above the rest
mass electron energy and thus allowing the trigger of the pair creation
process. The increase of the associated pair pressure is thus able to
disrupt the plasma flow and eventually, for too high a pressure (due to
catastrophic run-away pair production for example) to smooth it
completely. Reversely, significant changes of the flow velocity profile
may modify the distribution function of the accelerated particles in such
a way that the number of particles enable to trigger the pair creation
process can change, modifying
consequently the pair creation rate.\\

This non-linearity of the problem is in addition to the intrinsic non
linearity of the pair creation process, which may be more or less
important depending on the compactness of the soft and hard photons field
in the source (Svensson \cite{sve87}). The pair creation process is also
naturally non-local in the sense that pairs are produced by photons which
are themselves emitted by particles located in an other part of the
plasma. The complete 3D--treatment of this problem is out of the scope of
this paper. We propose however to use simplifying assumptions in
1D--geometry, to study the global behavior of a shock structure in an
environment dominated by pairs.\\

This paper is divided as follows. We first present in section 2 the basic
hypotheses we use to treat (relatively easily) the problem without loss
of generality. Some important definitions and descriptions of the
geometry of our model are detailed in section 3. We then present the main
equations describing the behavior of a shock in a pair environment in
section 4. In section 5, we first discuss the case without hydrodynamical
feedbacks on the pair process, deducing a limit of the pair creation rate
above which the shock disappears. When feedbacks are taken into account,
we study in section 6 the existence of stationary states where pair
creation and hydrodynamical effects balance. We discuss the results in
section 7 before concluding.\\
{ Throughout the paper upstream and downstream quantities are denoted by
subscripts 1 and 2 respectively. Particles and photons energies are
expressed (unless pointed out) in unit of $m_ec^2$, speeds in unit of
$u_1$, the upstream flow velocity, lengths in unit of the upstream
diffusion length $D_1/u_1$, where $D_1$ is the upstream diffusion
coefficient, and times in unit of $D_1/u_1^2$. Variables in reduced units
are overbraced by a $\sim$.}

\section{Basic hypotheses}
\label{basic}
\subsection{Simple geometry}
We assume a 1D geometry (the $x$ axis being align with the flow velocity
cf. Fig. \ref{mainchoc}) meaning that the different parameters
characterizing the flow are homogeneous in each section perpendicular to
the $x$ axis. However, we will use the transverse choc radius $R$ to
evaluate the size of the pair creation region $R_{\gamma\gamma}$ in
section 3.2.2. The 1-D assumption is certainly well justified in the
central part of the flow where boundary effects are negligible. We also
make the assumption of a parallel shock, i.e. the magnetic field lines
are supposed to be parallel to the normal of the shock front. We however
assume the magnetic field to be slightly perturbed near the shock and we
admit these perturbations to be dominated by Alfven waves. It is thus
possible for particles to be scattered by these waves through pitch angle
scattering (Jokipii \cite{jok76}; Lacombe \cite{lac77}) and thus going
back and forth across the shock.

\subsection{Constant coefficient diffusion}
Rigorously the upstream and downstream spatial diffusion coefficient,
${D_{1,xx}}$ and ${D_{2,xx}}$, depend slightly on
$\gamma$. For small Larmor radii of the particles compared to the large
scale of the turbulence (which is well--verified for leptons) one can
show (Casse et al.  \cite{cas01}) that $\displaystyle D_{i,xx}\propto
\gamma^{2-\beta}$ where $\beta=5/3$ for Kolmogorov spectrum. The
dependence on $\gamma$ being weak, we then consider $D_{i,xx}$ to be
independent of the Lorentz factor $\gamma$ of the particles. We also
assume $D_{i,xx}$ to be independent of $x$. Since we work in
1D--geometry, the upstream and downstream coefficient diffusion will be
simply noted $D_1$ and $D_2$ respectively in the following.

\subsection{Plasma dominated by the relativistic pressure}
We suppose the plasma pressure $P_{tot}$ to be dominated by the pressure
$P_{rel}$ of the relativistic leptons population. This seems relatively
reasonable since the leptons acceleration time scale is expected to be
relatively small (and largely smaller than the proton acceleration time
scale, Henri et al. \cite{hen99}). $P_{rel}$ becomes thus rapidly larger
than the thermal pressure. In this case, the flow velocity profile make
already a smooth transition between the up and downstream region but
acceleration still occurs (Drury et al. \cite{dru82}; Schneider \& Kirk
\cite{sch87}). Besides, with this assumption, the specific heat ratio of
the accelerated particle plasma is equal to 4/3. Consequently, in the
case of a strong shock, the compression ratio reaches the value of 7.



\subsection{Assuming a large Alfvenic Mach number}
{ Although the magnetic field is necessary to produce the stochastic
  acceleration, we will assume that the Alfven velocity
  $V_A=B/\sqrt{\rho}$ (where $B$ is the magnetic field amplitude and
  $\rho$ the plasma density) is much smaller than the upstream flow
  velocity $u_1$ i.e. the Alfvenic Mach number $M_{A}=u_1/V_A$ is very
  large. This leads to important simplifications in the treatment of the
  acceleration process (Vainio \& Schlikeiser \cite{vai98}, \cite{vai99};
  Schlickeiser et al. \cite{sch93}; Henri et al. \cite{hen99}):
\begin{itemize}
\item the second order Fermi process is negligible compared to the first
  order one since the ratio of the acceleration time scale
  is approximately $t_{1st}/t_{2nd}\simeq M_{A}^{-2}$ 
\item The scattering center compression ratio (which controls the
particle distribution spectral index) is close to the gaz compression
ratio $r=u_1/u_2$
\end{itemize}
Consequently, in the following, we will neglect the second order Fermi
process in our treatment and we will only use the gaz compression ratio
for the computation of the particle distribution function (cf. section
\ref{partdist}).  }

\subsection{Compton cooling in Thomson regime}
\label{threg}
{ We assume the particles cooling to be dominated by the Compton
  process onto the soft photons, neglecting the Synchrotron process.
  Again it is justified in the case of low magnetic energy density. For
  simplicity, the external soft photons field is supposed to be
  monoenergetic and we note $\epsilon_s$ the corresponding soft photon
  energy. This is not a bad approximation if the soft photons field is
  emitted by a thermal plasma like that probably associated with the Blue
  Bump emission seen in most AGNs. Furthermore we will restrict our study
  to the Thomson regime i.e.  $\epsilon_s\gamma_c <$1 where $\gamma_c$ is
  the high energy cut-off Lorentz factor of the particle distribution
  (cf. section \ref{highcut}).}

\begin{figure*}
\centering
\includegraphics[width=15cm]{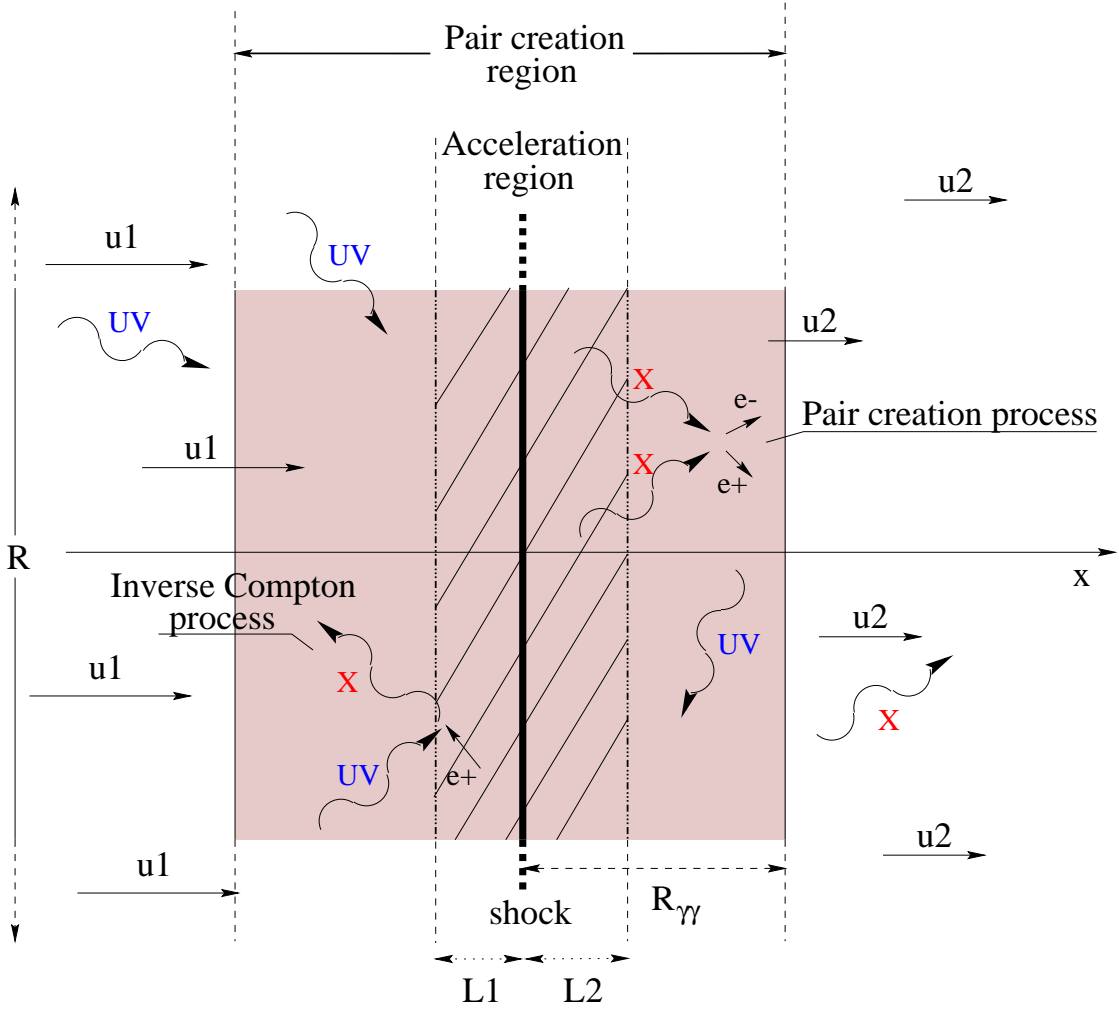}
\caption{Schematic view of a shock. The shock discontinuity is represented by
the vertical bold line.  We have also indicated the different parameters
defining the acceleration and pair creation region (cf. section 3). The
acceleration region is the region in between the 2 vertical dot-dashed
line. The pair creation one is filled in grey. Particles are represented
by straight arrows and photons by warped ones. Scales are not
respected. \label{mainchoc}}
\end{figure*}

\section{Some important parameters and definitions}

\subsection{The acceleration process}
\subsubsection{The acceleration threshold}
\label{accth}
Only particles having a Larmor radius $r_L$ comparable to a wavelength of
the Alfven spectrum will undergo scatterings (Jokipii \cite{jok76};
Lacombe \cite{lac77}) and thus will go back and forth many times across
the shock. However, in $e^--p^+$ plasma as those we deal with, the
non-relativistic protons limit the Alfven waves spectrum to wavelength
greater than $2\pi V_A/\omega_{cp}$ where $\omega_{cp}$ is the cyclotron
pulsation of the protons in a magnetic field B. Thus a particle of
momentum $p$ will be scattered by Alfven waves if it verifies:
\begin{equation}
  r_L=\frac{p}{|qB|}\geq \frac{m_pV_A}{|qB|}
  \label{reson}
\end{equation}
that is $p\geq m_pV_A$. This puts a severe threshold for resonant
interactions, especially for the leptons that must be already very
energetic to participate to the diffusion processes. Indeed, from
Eq. (\ref{reson}) we can deduce the lower Lorentz factor for a
relativistic lepton to be accelerated in a shock:
\begin{equation}
\gamma_{scat}=\frac{m_p}{m_e}\frac{V_A}{c}
\label{gscat}
\end{equation}
which could be easily of the order of a few. Several processes are known
to accelerate electrons up to such an energy as the development of a
parallel electric field component in magnetic reconnections, or waves
(magnetosonic or whistler { which have much lower particle energy
  threshold for resonant interaction}, Ragot \& Schlickeiser
\cite{rag98}). We will assume that such process accelerates the particles
up to a characteristic Lorentz factor $\gamma_{min}$ for which the
pre--acceleration and cooling times are equal. We further assume that
$\gamma_{min}$ is above the
scattering threshold $\gamma_{scat}$.\\

{ It is worth noting that the high energy power law spectra observed
  in compact objects generaly extend down to the keV range. For
  non-thermal models, it implies that the injection threshold is low
  ($\gamma_{scat}<$10), which is consistent with a small Alfvenic
  velocity.}

\subsubsection{The acceleration region}
\label{accreg}
Particles interacting with the shock are those located within about one
diffusion length of the shock, where the upstream and downstream
diffusion length are defined by (as previously said, i=1, 2 for up and
downstream respectively):
\begin{equation}
L_i= \frac{D_i}{u_i} 
\label{eqld}
\end{equation}
For a velocity profile distort by the presence of pairs, this definition
should not be rigorously used. However, we expect
the change of the velocity in the vicinity of the shock, due to the pair
pressure, to be relatively small, of the order of the unity (the
compression ratio $r=u_1/u_2$ can only vary between 1 and 7). The region
explored by particles will thus have a size not very different from that
given by Eq. (\ref{eqld})
We will call in the following ``acceleration'' region the physical space
defined by $-L_1\leq x\leq +L_2$ (cf. Fig. \ref{mainchoc}).

\subsubsection{The acceleration time scale}
\label{sectacc}
At each back and forth across the shock front, a relativistic particle
gain an energy such that:
\begin{displaymath}
\frac{\delta p}{p}=\frac{4}{3}\frac{u_1-u_2}{v\cos\theta_1}\nonumber
\end{displaymath}
where $v$ is the particle velocity and $\theta_1$ is the angle of the
magnetic field line with respect with the shock normal. In our case (a
parallel shock) we assume $\theta_1=0$.\\

On the other hand, the residence time $t_r$ of this particle in the
vicinity of the shock is determined by the diffusion coefficient:
\begin{displaymath}
{{t_r\simeq \frac{D_1}{u_1^2}+ \frac{D_2}{u_2^2}}}\nonumber
\end{displaymath}
{ If the upstream and downstream diffusion coefficients were equal, the
downstream contribution would be dominant. However, as shown by Vainio \&
Schlickeiser (\cite{vai98}), there is generally an increase of the
turbulence level in the downstream flow so that $D_1$ is expected to be
significantly larger than $D_2$. We refrain to use too much involved
calculations and we assume that $t_r$ is simply of the order of
the first term, i.e. $t_r\simeq \displaystyle\frac{D_1}{u_1^2}$.}\\

At each crossing, the particle has a probability
$\displaystyle\eta\simeq\frac{4u_2}{v}$ to escape (Bell, 1978) and for a
relativistic particle the crossing frequency is roughly $\displaystyle
\nu_c\simeq\frac{1}{\eta t_r}$. The acceleration rate is therefore (with
a factor of the order of the unity):
\begin{displaymath}
\frac{\langle\Delta p\rangle}{p\Delta t}\simeq\frac{r-1}{3t_r}.\nonumber
\end{displaymath}
We can thus deduce the corresponding acceleration time scale:
\begin{equation}
{{t_{acc}\simeq
3\frac{D_1}{u_1^2}\frac{1}{r-1}=\frac{D_1}{u_1^2}\wt{t}_{acc}}}
\label{TOA}
\end{equation}


\subsection{The pair creation process}
\subsubsection{The pair creation threshold}
\label{gammath}
A soft photon of energy $\epsilon_s$ (in reduced units) scattered by a
lepton of Lorentz factor $\gamma$ will be boosted, via IC process, to an
energy $\epsilon\simeq\displaystyle\frac{4}{3}\gamma^2\epsilon_s$. The
high energy photon produced will be able to give a pair of
electron-positron if at least :
\begin{equation}
\epsilon\gta 1 \mbox{ i.e. } \gamma\gta
\left(\frac{3}{4\epsilon_s}\right )^{1/2}=\gamma_{th}
\label{eqgth}
\end{equation}
A particle needs thus to have a Lorentz factor higher than $\gamma_{th}$
to generate a pair electron/positron. To fix an order of magnitude, for a
UV bump peaking near 10 eV (a common value in AGNs, Walter et al.
\cite{wal94}), Eq. (\ref{eqgth}) gives $\gamma_{th}\simeq$ 300.  We will
suppose that $\gamma_{min}\le\gamma_{th}$ so that particles need to be
accelerated in the shock to be able to trigger the pair creation process.

\subsubsection{The pair creation region}
\label{pairreg}
If the shock region is compact enough and if part of its radiation
extends beyond 511~keV, the pair creation, through photon-photon
annihilation, will become important.  The pair creation optical depth
$\tau_{\gamma\gamma}$ depends on photon energy $\epsilon$. Photons with
$\epsilon > 2$ produced in Compton scatterings will produce pairs mostly
with a Lorentz factor $\gamma\simeq\epsilon/2$ by colliding with photons
with energies of the order of 2$/\epsilon$. Consequently, the optical
depth for $\epsilon >$2 can be written as follows (Zdziarski \& Lightman
\cite{zdz85}):
\begin{eqnarray}
\tau_{\gamma\gamma}(\epsilon)&\simeq&\frac{2}{3}\sigma_T R
\frac{n_{ph}\left(\frac{2}{\epsilon}\right)}{\epsilon}\\
&=&\frac{R}{L_{\epsilon}}
\label{eqtau}
\end{eqnarray}
where $n_{ph}(\epsilon)$ is the density of photon with energy $\epsilon$
per unit volume and dimensionless energy $\epsilon$, and $\displaystyle
L_{\epsilon}=\left(\frac{2}{3}\sigma_T\frac{n_{ph}\left(\frac{2}{\epsilon}\right)}{\epsilon}\right)^{-1}$
is the typical attenuation length of a photon with an energy $\epsilon$.
Since we expect the photon density to decrease with energy,
$L_{\epsilon}$ is a decreasing function of $\epsilon$ and thus
$L_{\epsilon}<L_o$ where we define:
\begin{displaymath} 
L_o=L_{\epsilon =1}\nonumber
\end{displaymath}

On the other hand, the photon density will suffer from geometrical
dilution after some distance comparable to $R$, the transverse size of
the shock (cf. Fig. \ref{mainchoc}). Hence, we will define the pair
creation region as the region included between $-\Rgg$ and $+\Rgg$, where
$\Rgg$ is defined by:
\begin{equation}
 \Rgg=\frac{RL_o}{R+L_o}
\label{rggeq}
\end{equation}
This definition has the advantage to take into account both photon-photon
depletion and geometrical dilution effects. In the following, we will
suppose, for sake of simplicity, that {\it the pair creation rate is
constant in this region}.


\subsection{The flow profile and particle distribution}
\label{partdist}
\subsubsection{The ``effective'' compression ratio}
\label{reffec}
The definition of the compression ratio like the ratio between the far
upstream and downstream flow velocity is rather unsatisfactory in our
case since it will not give a real estimate of the velocity change
experienced by a relativistic particle in the vicinity of the shock,
where the particle distribution is principally built. We will thus define
an ``effective'' compression ratio, noted simply $r$, as the ratio
between the upstream velocity in $-10L_1$ and the downstream velocity in
$+10L_2$. We have checked that in the case of a strong shock without
pairs, this definition still gives a compression ratio very near the
expected value of 7.

\subsubsection{The spectral index}
It can be shown that the solution of the evolution equation of the
particle distribution function including the pair creation process (cf.
Eq. (\ref{eqfb})), still has a energy power law dependence, as it is
effectively the case without pairs (cf. Appendix \ref{app1}). The
spectral index $s$ \footnote{here $s$ is the spectral index of the
  spatially integrated distribution function of the particles
  $n(\gamma)\propto\gamma^{-s}$. It is related to the spectral index $q$
  defined in Appendix by $s=q-2$} keeps also the same expression in
function of the compression ratio i.e.:
\begin{equation}
 s(r)=q-2=(r+2)/(r-1).
\label{eqs}
\end{equation}
For plasma dominated by relativistic pressure, the compression ratio is
necessarily smaller than 7 so that $s(r)$ is larger than 1.5.

In fact, Eq. (\ref{eqs}) is really valid for not too large a relativistic
pressure in comparison to the thermal one ($P_{rel}/P_{th}$ must be
smaller than $\sim$ 10, cf. Pelletier \& Roland \cite{pel84}) and we
assume to be in this limit case so that we can also still neglect the
thermal pressure of the plasma in comparison to the relativistic one.

\subsubsection{The high energy cut-off}
\label{highcut}
Since we take into accounts the cooling, a high energy cut-off in the
particle distribution must necessarily appear at a Lorentz factor
$\gamma_c$ where heating and cooling balance (Webb et al.,
\cite{web84}). Since we suppose that particles cool via inverse Compton
process, and assuming that the soft photon density is homogeneous in the
shock region, the mean cooling time scale can be written:
\begin{equation}
 t_{cool} =\frac{3}{4}\frac{m_{e}c}{\gamma \sigma_T W_{soft}}=
 \frac{3}{4}\frac{1}{l_s\gamma}\frac{u_1}{c}\wt{R}\frac{D_1}{u_1^2}
\label{TOAb}
\end{equation}
where $l_s$ is the local soft compactness $\displaystyle
l_s=\frac{L_{soft}\sigma_T}{4\pi Rm_ec^3}$, $m_e$ being the electron
mass,$L_{soft}$ the soft radiation luminosity injected into the shock
region and $W_{soft}$ the corresponding local soft photon energy density.
The maximum Lorentz factor $\gamma_c$ achievable by the acceleration
process is thus obtained by setting $t_{acc} = t_{cool}$ which gives,
using Eqs.  (\ref{TOA}) and (\ref{TOAb}):
\begin{equation}
{\gamma_c=\frac{\wt{R}}{l_s}\frac{u_1}{c}\frac{r-1}{4}}
\label{GAM}
\end{equation}


\section{Basic kinetic equations at shocks}
The shock region is depicted in Fig. \ref{mainchoc} with the acceleration
and pair creation region described above. The scale are not respected. We
take the origin of the x-axis at the point where the second derivative of
the flow velocity vanishes, that is:\\
\begin{equation}
  \left.\frac{\partial^2 \wt{u}}{\partial
  \wt{x}^2}\right|_{\wt{x}=0}=0.  \label{chocen0}
\end{equation}

As previously said, we suppose the existence of a magnetic field B,
slightly perturbed by Alfven waves. The particles are thus scattered by
these waves trough pitch angle scattering (Jokipii \cite{jok76};
Lacombe \cite{lac77}) and can cross the shock front several times before
escaping unless they are rapidly cooled by radiative processes.  During
these scatterings, the particle gain energy trough the well known first
order Fermi process. We also suppose the magnetic perturbations to have
sufficiently small amplitudes so that we can treat the problem in
quasilinear theory using the Fokker-Planck formalism. Besides we assume
that, in each part of the shock front, the scattering is sufficient for
the particle distribution to be nearly isotropic. With these different
assumptions, and when first order Fermi process just as radiative losses
and pair creation/annihilation are taken into account, the particles
distribution function $f(p,x)$ must verify the following equation :
\begin{eqnarray}
  \dt{f}+u\dx{f} &=& \frac{1}{3}\dx{u} p
  \dpp{f}+\frac{1}{p^2}\dpp{bp^4f}+\dx{}D\dx{f} \nonumber\\ & & +\A+\B+\C.
  \label{eqfb}
\end{eqnarray}
The three first terms of the right member correspond to the first order
process, the radiative losses ($b$ being $>$ 0, { in the case of
  Compton cooling in the Thomson regime} $b=1/(\gamma t_{cool})$ where
$t_{cool}$ is given by Eq.  \ref{TOAb}) and the spatial diffusion
respectively, $\A$ is the injection of particles at $\gamma_{min}$ due to
the pre-acceleration processes (cf.  section \ref{accth}), $\B$ is the
pair creation rate and $\C$ the annihilation one.  This equation is
essentially the equation of the cosmic-ray transport originally given by
Parker (\cite{par65}), Skilling (\cite{ski75} and references therein)
except for the addition of the radiative, pre-acceleration and pair
processes.

\subsection{Evolution equation of the flow velocity profile}
\label{secpair}
As previously said in section \ref{basic}, we supposed the shocked plasma
pressure to be dominated by the pressure of the relativistic particles.
The latter can thus be written as follows:
\begin{displaymath}
P_{\mbox{{rel}}}=\int 4\pi p^2 \frac{pc}{3} f(p,x)dp.\nonumber
\end{displaymath}
Consequently, by multiplying Eq. (\ref{eqfb}) by $4\pi p^3c/3$ and
integrating on $p$ we obtain the hydrodynamic equation linking pairs
(through their pressure) and the flow velocity that is (in stationnary
state):
\begin{equation}
  u\dx{P_{\mbox{{rel}}}}+\frac{4}{3}P_{\mbox{{rel}}}\dx{u}
=\dx{}D \dx{}P_{\mbox{{rel}}}+
\dot{Q}^-+\dot{P}_{\pm}+\dot{Q}^+
\label{equadiffPpaire}
\end{equation}
where:
\begin{displaymath}
\dot{Q}^-=\int 4\pi\frac{pc}{3}\dpp{bp^4f}dp\nonumber
\end{displaymath}
corresponds to the pressure loss rate due to radiative losses (it is thus
obviously negative),
\begin{displaymath}
\dot{Q}^+=\int\frac{4\pi p^3c}{3}\A dp\nonumber
\end{displaymath}
is the pair pressure creation rate due to the pre-accelerator processes, and
\begin{displaymath}
\dot{P}_{\pm}=\int\frac{4\pi p^3c}{3}\B dp+\int\frac{4\pi p^3c}{3}\C
dp\nonumber
\end{displaymath}
is the pressure creation rate including the pair creation and
annihilation processes.
It is now possible to deduce from Eq. (\ref{equadiffPpaire}) the
differential equation followed by the flow velocity. Indeed the momentum
conservation equation gives:
\begin{displaymath}
  \rho\left(\dt{u}+u\dx{u}\right)+\dx{P_{\mbox{rel}}}=0.\nonumber
\end{displaymath}
which, in stationnary state, is easily integrated:
\begin{equation}
  \rho u^2 + P_{\mbox{rel}} = \cal{C}
  \label{equamouv}
\end{equation}
where $\cal{C}$ is a constant. Assuming the mass is dominated by protons
(i.e. $\rho=n_pm_p$) and thus is conserved, Eq. (\ref{equamouv}) gives:
\begin{displaymath}
  P_{\mbox{rel}}=\rho_1 u_1(u_1-u).\nonumber
\end{displaymath}
To obtain this equation we have supposed
$\left. P_{\mbox{{rel}}}\right|_{-\infty}=0$ and
$\left. u\right|_{-\infty}=u_1$. Then Eq. (\ref{equadiffPpaire}) gives
finally:
\begin{equation}
  \frac{4}{3}\dtx{u}-\frac{7}{3}\wt{u}\dtx{u}+\frac{\partial^2}{\partial
  \wt{x}^2}\wt{u} = \dttx{} \left(\frac{4}{3}\wt{u} -\frac{7}{6}
  {\wt{u}}^2 + \dtx{u}\right)= \frac{\Pi}{\wtRgg}
  \label{equadiffu}
\end{equation}
where the reduced variables are defined as followed:
\begin{eqnarray}
  \wt{u} &=& \frac{u}{u_1}\\
  \wt{x} &=& \frac{x}{L_1} \mbox{ and } \wtRgg = \frac{\Rgg}{L_1}\\
  \Pi &=&
  \frac{\dot{Q}^++\dot{P}_{\pm}+\dot{Q}^-}{\rho_1
  u_1^3}\Rgg \label{A}.
\end{eqnarray}
\begin{figure}[t]
\includegraphics[width=\columnwidth]{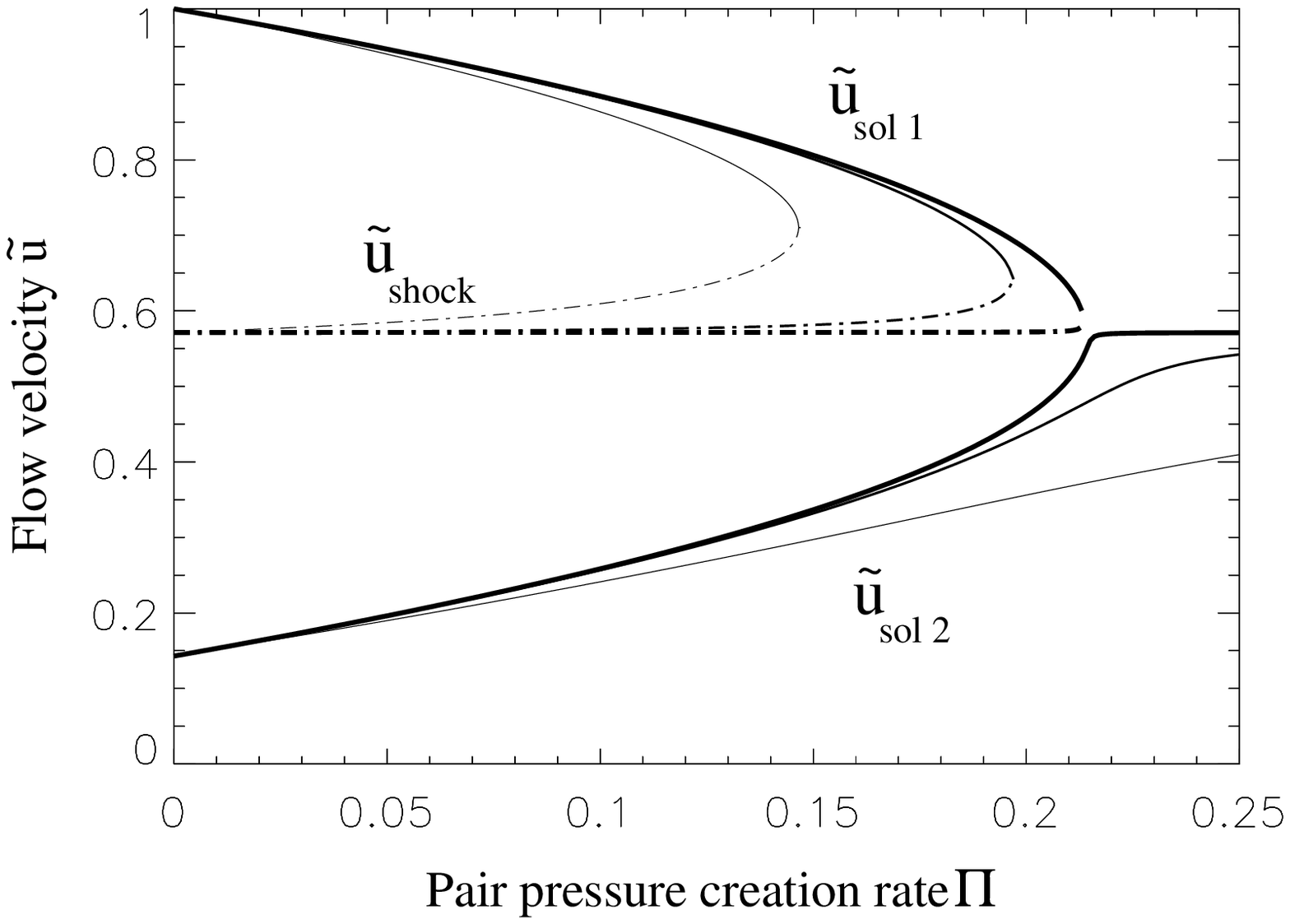}
\caption{Plot of the 3 solutions of Eq. (\ref{pu}) versus $\Pi$ for
different value of $\wtRgg$. From thin to thick lines, $\wtRgg$= 10, 100,
10$^3$. The flow velocity $\widetilde{u}_{\rm{shock}}$ at the shock
location, i.e. in $\widetilde{x}=0$ is plotted in dashed line.  For each
value of $\wtRgg$, there is a maximal value of $\Pi$, above which only
one real solution exists. The disapperance of the
$\widetilde{u}_{\rm{shock}}$ branch means that the shock cannot exist
anymore.
\label{a_urelation}}
\end{figure}
The parameter $\Pi$ is the ratio of the flux of energy transmitted to
particles, through pair creation/annihilation, pre-acceleration and
cooling processes, integrated on the ``1D'' volume $\wtRgg$, to the flux
of kinetic energy of the upstream flow. As the acceleration and the
cooling time are short with respect to the travel time $\Rgg/c$, we can
assume that the sum of the 3 terms $\dot{Q}^+$, $\dot{P}_{\pm}$, and
$\dot{Q}^-$ of Eq. (\ref{A}) is equivalent to the instantaneous creation of
pairs at the Lorentz factor $\gamma_{min}$, that is:
\begin{eqnarray}
\dot{Q}^++\dot{P}_{\pm}+\dot{Q}^-&\simeq&\frac{\gamma_{min}m_ec^2}{3}\int
4\pi p^2\B dp\nonumber\\
&=&\frac{\gamma_{min}m_ec^2}{3}\dot{n}_{\pm}\label{AA}
\end{eqnarray}
where $\dot{n}_{\pm}$ is the total (integrated over the particle Lorentz
factor) pair creation rate.  Then, $\Pi$ can also be seen as the ratio of
the pair luminosity (the pair power density integrated on the ``1D''
volume $\Rgg$) to the kinetic energy flux of the flow. This parameter
will play an important role in the evolution of the shock profile as we
will see in the following. We can anticipate that a value of $\Pi$ of the
order of the unity will be certainly unfavorable to the formation of the
shock. It means that all the kinetic energy of the upstream flow will be
dissipated in radiative cooling and pair creation/annihilation processes.

\section{Solution without feedback}
\label{chocdes}
In this section, we study the solutions of the previous equations as a
function of the parameter $\Pi$. We do not consider the feedback of
the hydrodynamics on the pair creation process, i.e. $\Pi$ is
considered as a fixed free parameter. The complete problem, which must
take into account the different processes to find self consistent
solutions, will be treated in the next section.

\subsection{A maximal pair creation rate}
\label{maxpair}
A way to solve Eq. (\ref{equadiffu}) is to integrate it between $-\infty$
and $L_2$. In this case, we can neglect the cooling during the
integration. Indeed, even if for $x\le -L_1$ particles do not interact
with the shock, we have assumed that some processes apply and balance
coolings so that particles are injected at a minimum Lorentz factor
$\gamma_{min}$ above the resonance threshold $\gamma_{scat}$. On the other
hand, in the shock region (that is $-L_1\le x\le L_2$), we have seen in
section \ref{highcut} that the
\begin{figure*}
\centering
\includegraphics[width=14cm]{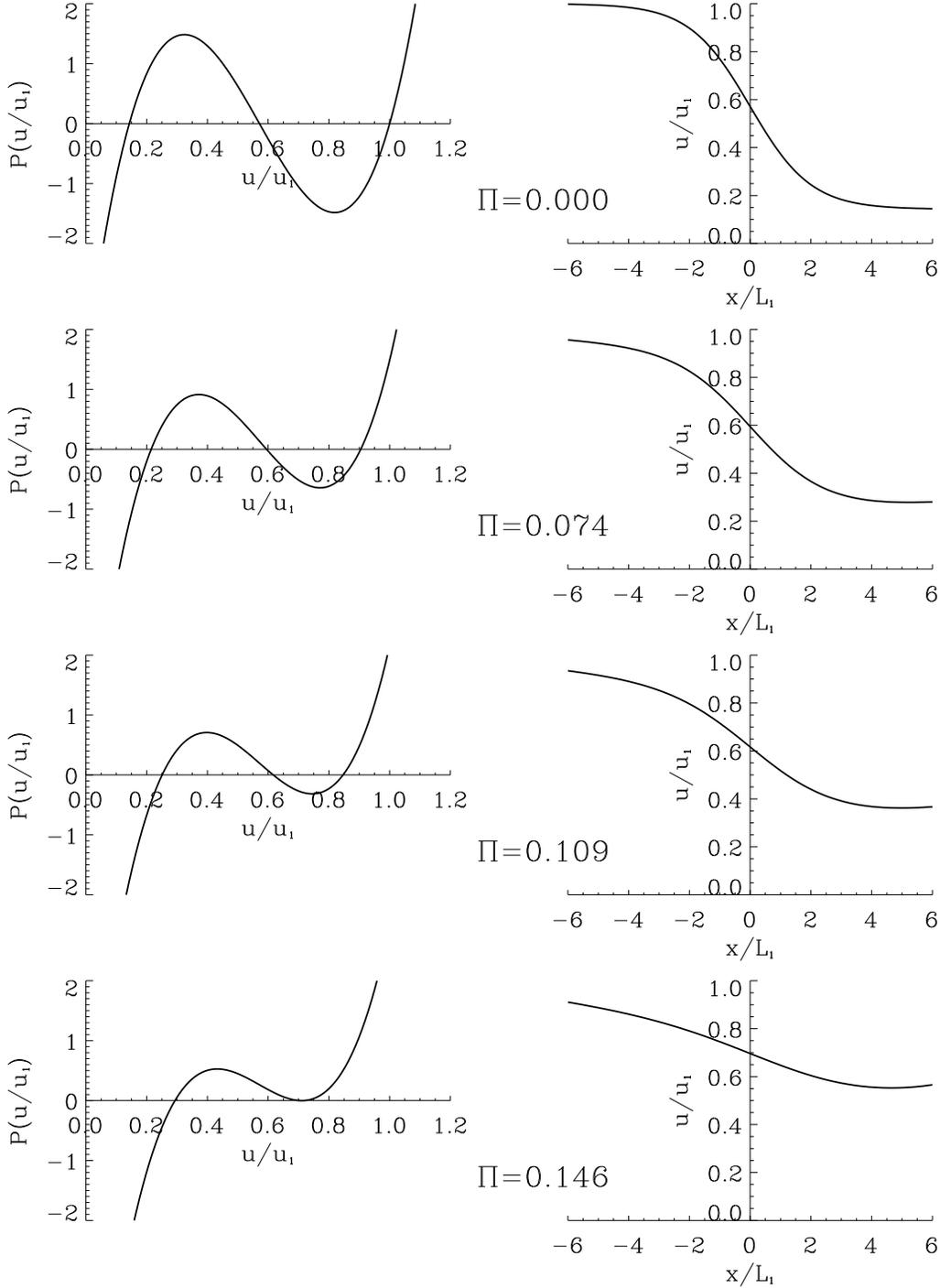}
\caption{disappearance of a shock due to pair creation. We have taken
$\wtRgg$=10 meaning that $\Pi$ must be smaller than $\sim 0.15$
(cf. Eq. \ref{condnec}).  \label{profchoc}}
\end{figure*}
coolings are negligible, in comparison to heatings, for particles with a
Lorentz factor smaller than $\gamma_c$ (the case of the majority of the
particles). Besides, we can also neglect the annihilation since it occurs
mainly at low energy, i.e. for particles with Lorentz factor
$\gamma\simeq 1$ (cf. Coppi \& Blandford \cite{cop90}), whereas we
assume $\gamma\ge\gamma_{min}>1$.  In these conditions, and with the
assumption that the pair creation is homogeneous in the region
$-\Rgg<x<+\Rgg$ (meaning $\dot{P}_{\pm}$ roughly constant) and null
outside (meaning $\dot{P}_{\pm}$=0), the integration of Eq.
(\ref{equadiffu}) gives:
\begin{equation}
  \dtx{u}=\frac{7}{6}(1-\wt{u})\left(\frac{1}{7}-\wt{u}\right)
  +\Pi\left(\frac{\wt{x}}{\wtRgg}+1\right) \label{equadiffub}.
\end{equation}
Equation (\ref{equadiffub}) can be solved with the following boundary
conditions:
\begin{displaymath}
  \wt{u}(-\wtRgg) = \wt{u}^*(-\wtRgg), \nonumber\\
\end{displaymath}
where $\wt{u}^*$ is the flow velocity solution of Eq. (\ref{equadiffub})
when $\Pi=0$ that is (Drury et al. \cite{dru82}):
\begin{displaymath}
  \wt{u}^*(x)=\frac{4}{7}-\frac{3}{7}\tanh\left[
    \frac{2}{7}\int_{x_0}^{x}\frac{dx'}{D}\right],\nonumber
\end{displaymath}
and assuming Eq. (\ref{chocen0}). Using Eq. (\ref{equadiffub}), this last
equation becomes:
\begin{eqnarray}
  \left.\frac{\partial^2 \wt{u}}{\partial
      \wt{x}^2}\right|_{\tin{x=0}}=P(\wt{u})
      &=&49\wt{u}^3-84\wt{u}^2+(39+42\Pi)\wt{u}\nonumber\\ 
        & & -6\Pi\left(4+\frac{3}{\wtRgg}\right)-4=0 \label{pu}
\end{eqnarray}
This equation possesses in general 3 real solutions which obviously
depend on $\Pi$ and $\wtRgg$. They have been plotted in Fig.
\ref{a_urelation} as a function of $\Pi$ and for different values of
$\wtRgg$. Of course, if $\Pi$ vanishes the three branches of solution
converge respectively to the well-known results u=1/7, 4/7 and 1
corresponding to the values without pair creation. By continuity, the
flow velocity at the shock location will follow the second branch noted
$u_{\mbox{shock}}$ on the figure (plotted in dashed line). It appears
that, for a given value of $\wtRgg$, there exists a maximal value
$\Pi_{max}$ of $\Pi$ above which there is only one real solution which
still verifies Eq. (\ref{pu}). The unphysical discontinuity of the flow
velocity at the shock location for $\Pi=\Pi_{max}$ means simply that the
shock cannot exist anymore.\\

\begin{figure}
\includegraphics[width=\columnwidth]{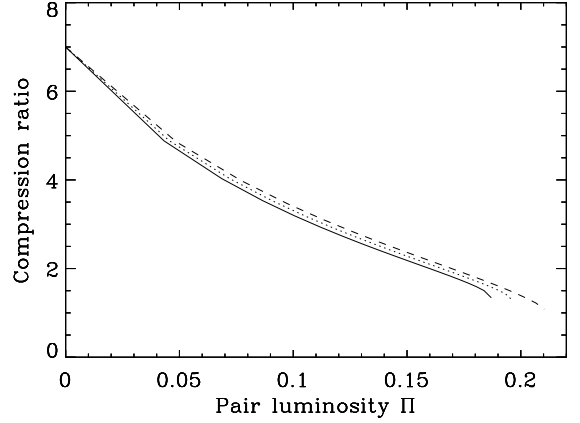} 
\caption{Variation of the compression ratio $r$ in function of the pair
  luminosity $\Pi$ for different values of $\wtRgg$. $\wtRgg$=10, 100,
  1000 for the solid, dotted and dot-dashed line respectively.
  \label{rfuna}}
\end{figure}

We have reported in Fig. \ref{profchoc}, different shape of the
polynomial $P(\wt{u})$ and the corresponding flow velocity profiles
obtained numerically by solving Eq. (\ref{equadiffub}) for different
values of $\Pi$ (fixing $\wtRgg$ to 10). We clearly see the softening of
the flow velocity profile when $\Pi$ increases meaning that the
acceleration becomes less and less efficient. We have also plotted in
Fig.  \ref{rfuna}, the variation of the compression ratio $r$ (as defined
in section \ref{reffec}) in function of the pair creation luminosity
$\Pi$ and for different values of $\wtRgg$. As expected, $r$ converge to
$\sim$~1 when $\Pi$ increases.\\

\begin{figure}[h]
\includegraphics[width=\columnwidth]{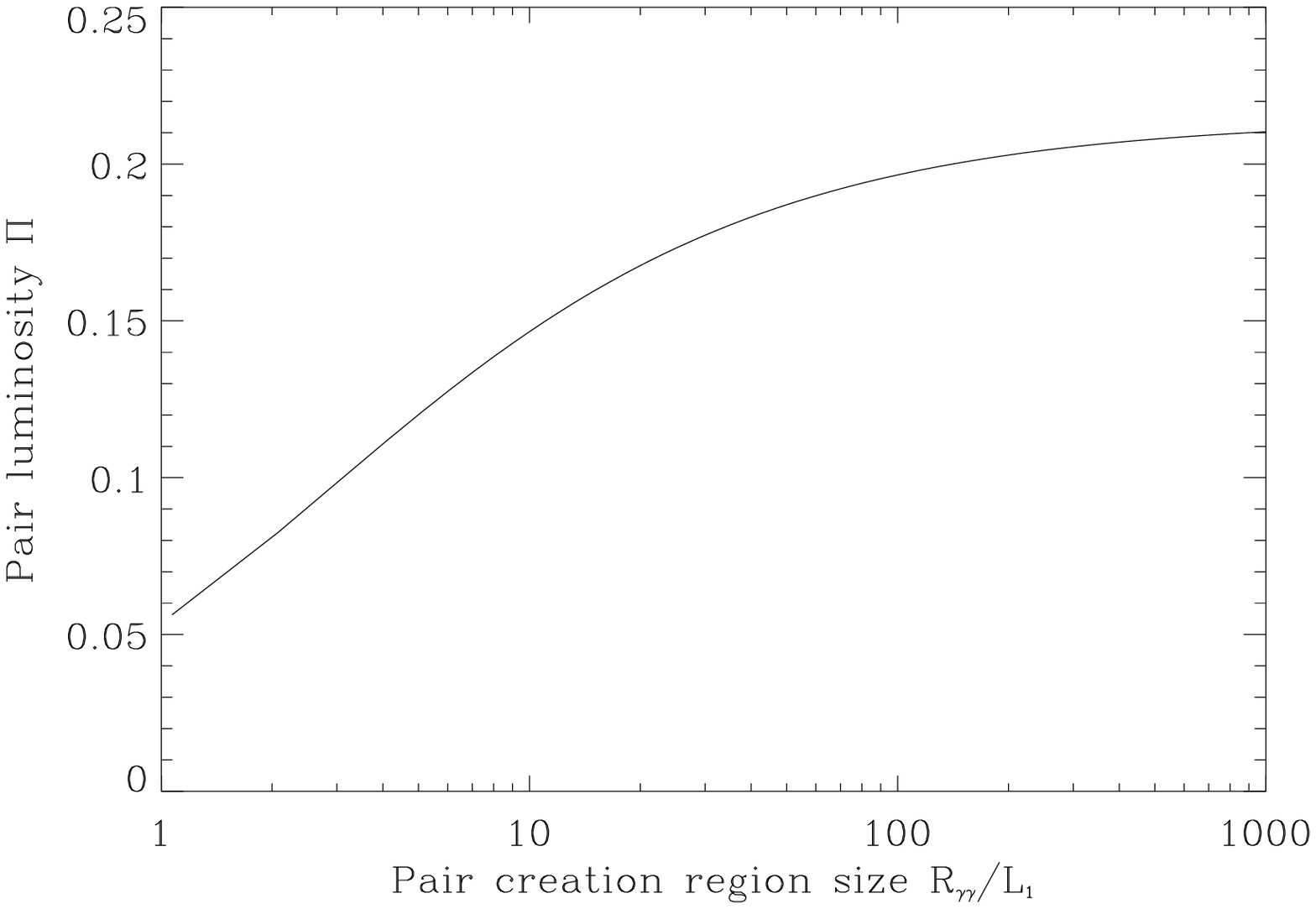}
\caption{Plot of the maximal pair luminosity $\Pi_{max}$ versus the size
  $\wtRgg=\Rgg/L_1$ of the pair creation region.
\label{a_lrelation}}
\end{figure}

Since $P(\wt{u})$ is a third degree polynomial, the transition between 3
to 1 real solution happens when the following conditions are satisfied:
\begin{eqnarray*}
  P(u) &=& 0\nonumber\\
  P'(u) &=& 0\nonumber
\end{eqnarray*}
The resolution of this system of equations gives thus a relation between
$\Pi_{max}$ and $\wtRgg$:
\begin{equation}
  \Pi_{max}=\frac{3}{14}-
  \frac{1}{14}\left(\frac{63\Pi_{max}}{\wtRgg}\right)^{2/3} \label{condnec}
\end{equation}
We have plotted $\Pi_{max}$ versus $\wtRgg$ in Fig.  \ref{a_lrelation}.
We see that it is an increasing function of $\wtRgg$, meaning that the
larger the pair pressure creation region the larger the
power we need to kill the shock.\\

But the important conclusion of this part is that $\Pi_{max}$ is
necessarily smaller than 3/14$\simeq$ 0.20 (cf. Eq. (\ref{condnec}))
meaning that {\it at most 20\% of the kinetic energy flux of the upstream
  flow transformed in pairs is sufficient to suppress the shock
  discontinuity}.\\

\section{Stationary states}
\label{feedback}
\subsection{The equations}
At this stage, we have studied the deformation of the flow profile due to
the pair creation as a function of the parameter $\Pi$. Conversely the
velocity profile of the flow will control the distribution function of
the accelerated particles, and hence the number of particles able to
trigger the pair creation process (i.e.  particles with
$\gamma\ge\gamma_{th}$). This back reaction will thus select the control
parameter $\Pi$ defining the possible stationary states for a given set
of external parameters.\\ 

From the previous section we have seen that the compression ratio $r$
decreases when the pair creation rate raises (cf. Fig. \ref{rfuna}) but
reversely if $r$ decreases, the spectral index will increase and the
number of high energy particles will decrease (cf. Eqs. (\ref{eqs}) and
(\ref{GAM})), the final effect being a decrease of the pair creation
rate.  Consequently, we indeed expect the system to reach, in some
conditions, stationary states where hydrodynamics and pair creation
effects balance.\\

The problem is now to calculate the pair production rate as a function of
the flow profile parameters. { For this goal, we first approximate the
downstream particle distribution function {\it just after the shock} with
the following form} :
\begin{equation}
n_{{2}}(\gamma{,x=0^+})= n_o\gamma^{-s}\exp\left
(-\frac{\gamma}{\gamma_c}\right ),
\label{distfunc}
\end{equation}
From the remarks of section \ref{partdist}, it appears to be a relatively
good approximation in the vicinity of shocks where first order Fermi
process and Compton coolings occur. The assumption of an exponential high
energy cut-off appears also to be reasonnable (cf. for exemple Webb et
al. \cite{web84}).  \\

The particle energy distribution is then completely determined by
$n_{0}$, $s$, and $\gamma_{c}$. The latter are function only of the
compression ratio $r$ and the external and geometrical parameters
$\wt{R}$, $l_{s}$ and $u_{1}/c$. We can thus express $n_{{2}}(\gamma{,x=0^+})$
as a function only of these external parameters and the two dimensionless
variables $r$ and
\begin{equation}
\label{eqwtno}
    \wtno=\frac {m_e}{\rho_1}n_{0}
\end{equation}
The compression ratio $r$ is a function of $\Pi$ and $\wtRgg$ through the
numerical integration of Eq. (\ref{equadiffub}). We can express also
$\wtno$ as a function of these two quantities in the following way. For
simplicity, we will neglect, in the acceleration processes, particles
coming from the upstream flow and consider the plasma to be dominated by
created pairs.  Consequently, assuming that the pair creation rate
$\dot{n}_{\pm}$ is constant within the pair creation region, the total
downstream particle density $n_{tot}$ after the shock can be written:
\begin{equation}
n_{tot} = \int_{\gamma_{min}}^{\infty}n_2(\gamma,x=0^+)d\gamma \simeq
\dot{n}_{\pm}\frac{\Rgg}{u_2}.\nonumber
\label{eqntot}
\end{equation} 
With the help of Eqs. (\ref{distfunc}) and (\ref{eqwtno}), we thus find:
\begin{equation}
\widetilde{n}_o\simeq\frac{m_e}{\rho_1}\dot{n}_{\pm}\frac{\Rgg}{u_2}
\frac{(s-1)}{\gamma_{min}^{1-s}}
\label{eqno}
\end{equation}
where we have use the approximation
$\displaystyle\int_{\gamma_{min}}^{\infty}\gamma^{-s}exp\left(-\frac{\gamma}{\gamma_c}\right)
d\gamma\simeq\frac{\gamma_{min}^{1-s}}{s-1}$, reasonable for $s>1$, which
is always our case (cf.
section \ref{partdist}).\\

As we assume that the produced pairs are instantaneously accelerated to
the Lorentz factor $\gamma_{min}$, the pair luminosity $\Pi$ can be
approximated by (cf. Eq. (\ref{A}) and (\ref{AA})):
\begin{equation}
\Pi\simeq\frac{(\gamma_{min}m_ec^2\dot{n}_{\pm})\Rgg}{3\rho_1u_1^3}.
\label{pcrate}
\end{equation}
Consequently, Eq. (\ref{eqno}) can be re-written:
\begin{equation}
\wtno=\Pi\frac{u_1^2}{c^2}\wtRgg
\frac{(s-1)}{\gamma_{min}^{2-s}}r
\label{eqnofin}
\end{equation}
which links $\wtno$ to $\Pi$ and $\wtRgg$.\\

To close our systems of equations, we now need to link the parameters
$\Pi$ and $\wtRgg$ to $\wtno$ and $r$ using two other independent
equations. This could be done by computing directly the pair creation
rate, as follows. The particles accelerated in the shock cool via IC
process on the surrounding external soft photons. The photon spectrum
emitted by a single particle of Lorentz factor $\gamma$ per unit of time
and dimensionless energy $\epsilon$ can be approximated by:
\begin{displaymath}
\left.\frac{dN_{ph}}{dtd\epsilon}\right|_{\gamma}=\frac{c}{R}\frac{l_s}{\epsilon_s}
\frac{1}{\sigma_T}\frac{d\sigma}{d\epsilon}
=\frac{c}{R}\frac{l_s}{\epsilon_s}
\delta(\epsilon-\frac{4}{3}\gamma^2\epsilon_s)\nonumber
\end{displaymath}
where we have approximated the differential Compton diffusion cross
section by a Dirac distribution, i.e
$\displaystyle\frac{d\sigma}{d\epsilon}=\sigma_T\delta
(\epsilon-\frac{4}{3}\gamma^2\epsilon_s)$.\\
The photon spectrum emitted by a particle of initial Lorentz factor
$\gamma$ cooling down to $\gamma=1$, is then given by:
\begin{equation}
\left.\frac{dN_{ph}}{d\epsilon}\right|_{\gamma}=\int_{\gamma}^{1}
\frac{dN_{ph}}{dtd\epsilon}(\gamma')
\frac{dt}{d\gamma'}d\gamma'
\label{eqnphdeps}
\end{equation}
where the cooling rate
$\displaystyle\frac{dt}{d\gamma'}=\frac{t_{cool}}{\gamma'}$, with
$t_{cool}$ given by Eq.  (\ref{TOAb}). The integration of Eq.
(\ref{eqnphdeps}) then gives:
\begin{displaymath}
\left.\frac{dN_{ph}}{d\epsilon}\right|_{\gamma}=
\left(\frac{3}{16\epsilon_s}\right)^{\frac{1}{2}}
\epsilon^{-\frac{3}{2}}[1-H(\frac{4}{3}\gamma^2\epsilon_s)]\nonumber
\end{displaymath}
where $H$ is the Heaviside step function.
It is now possible to compute the total photon spectrum emitted by the
distribution of the particles crossing the shock per unit of time and
dimensionless energy $\epsilon$:
\begin{displaymath}
\frac{dN^{tot}_{ph}}{dtd\epsilon}(\epsilon)=
\int^{\infty}_{\sqrt{\frac{3\epsilon}{4\epsilon_s}}}\dot{N}_{part}(\gamma)
\left.\frac{dN_{ph}}{d\epsilon}\right|_{\gamma}d\gamma
\end{displaymath}
where
$\displaystyle{{\dot{N}_{part}(\gamma)=n_2(\gamma,x=0^+)u_2\pi
R^2}}$ is the flux of particle crossing the shock. Consequently:
\begin{eqnarray}
\frac{dN^{tot}_{ph}}{dtd\epsilon}(\epsilon)&\simeq&\frac{\wtno
  l_{kin}}{r}\frac{c^2}{u_1^2}\frac{Rc}{\sigma_T}
\left(\frac{3}{16\epsilon_s}\right)^{\frac{1}{2}} 
\epsilon^{-\frac{s+2}{2}}\frac{\gamma_{th}^{1-s}}{s-1}\nonumber\\
& &\times\exp\left(-\frac{\gamma_{th}\sqrt{\epsilon}}{\gamma_c}\right)
\label{eqspecph}
\end{eqnarray}
where we define the kinetic compactness $\displaystyle
l_{kin}=\frac{(\rho_1 u_1^3\pi R^2)\sigma_T}{Rm_ec^3}$. We can then
deduce the pair creation rate due to the photon-photon annihilation on a
length scale $\Rgg$:
\begin{equation}
\dot{n}_{\pm}=\frac{1}{\pi R^2\Rgg}\int_{\epsilon=1}^{\infty}
\frac{dN^{tot}_{ph}}{dtd\epsilon}(\epsilon)\left(1-e^{-\tau_{\gamma\gamma}(\epsilon)}
\right)d\epsilon\label{eqX}
\end{equation}
where $\tgg(\epsilon)$ is given by Eq. (\ref{eqtau}). We finally obtain a
new relation between $\Pi$, $\wtno$ and $r$:
\begin{eqnarray}
\Pi&=&\frac{\gamma_{min}\wtno}{4r}\frac{c^2}{u_1^2}
\frac{\gamma_{th}^{2-s}}{s-1}\nonumber\\
& &\times\int\epsilon^{-\frac{s+2}{2}}
\exp\left(-\frac{\gamma_{th}\sqrt{\epsilon}}{\gamma_c}\right)
\left(1-e^{-\tau_{\gamma\gamma}(\epsilon)}\right)d\epsilon
\label{eqpifin}
\end{eqnarray}
Finally, $\wtRgg$ is given by Eq. (\ref{rggeq}) where:
\begin{equation}
L_o=\frac{3}{2\sigma_T n_{ph}(\epsilon=2)}
\end{equation}
and $n_{ph}(\epsilon)=\displaystyle\frac{1}{\pi
  R^2c}\frac{dN^{tot}_{ph}}{dtd\epsilon}(\epsilon)$.  Together with Eqs.
(\ref{equadiffub}), (\ref{eqnofin}) and (\ref{eqpifin}), this forms a
system of four equations of four unknowns $\Pi$, $\wtRgg$, $\wtno$ and
$r$ which has to be verified at the equilibrium i.e. when pair pressure
and hydrodynamics effects balance.\\
 
\subsection{The solutions}
\label{solutions}
The above system depends on six different parameters: $\gamma_{min}$, the
minimal Lorentz factor for the particles to be accelerated in the shock,
$\epsilon_s$, the soft photon energy (in unit of $m_ec^2$), $u_1$, the
upstream flow velocity, $\wt{R}$ the transverse size of the shock (in
unit of the diffusion length $L_1$), $l_s$, the
soft compactness and $l_{kin}$ the kinetic compactness.\\

The system is solvable in only some part of the parameter space. It
always possesses two solutions: a ``high pair density'' one (large
$\wtno$, large $\Pi$, small $r$) and a ``low pair density'' one (small
$\wtno$, small $\Pi$, large $r$). The latter connects to the trivial
solution of the problem i.e. $\Pi =0$. However, only the high pair
density states are compatible with the neglecting of particles coming
from the upstream flow (cf. Eq. \ref{eqntot}). Thus we will only focus,
in the following, on this branch of solutions.\\

\begin{center}
\begin{figure*}
\begin{tabular}{cc}
\includegraphics[width=0.95\columnwidth]{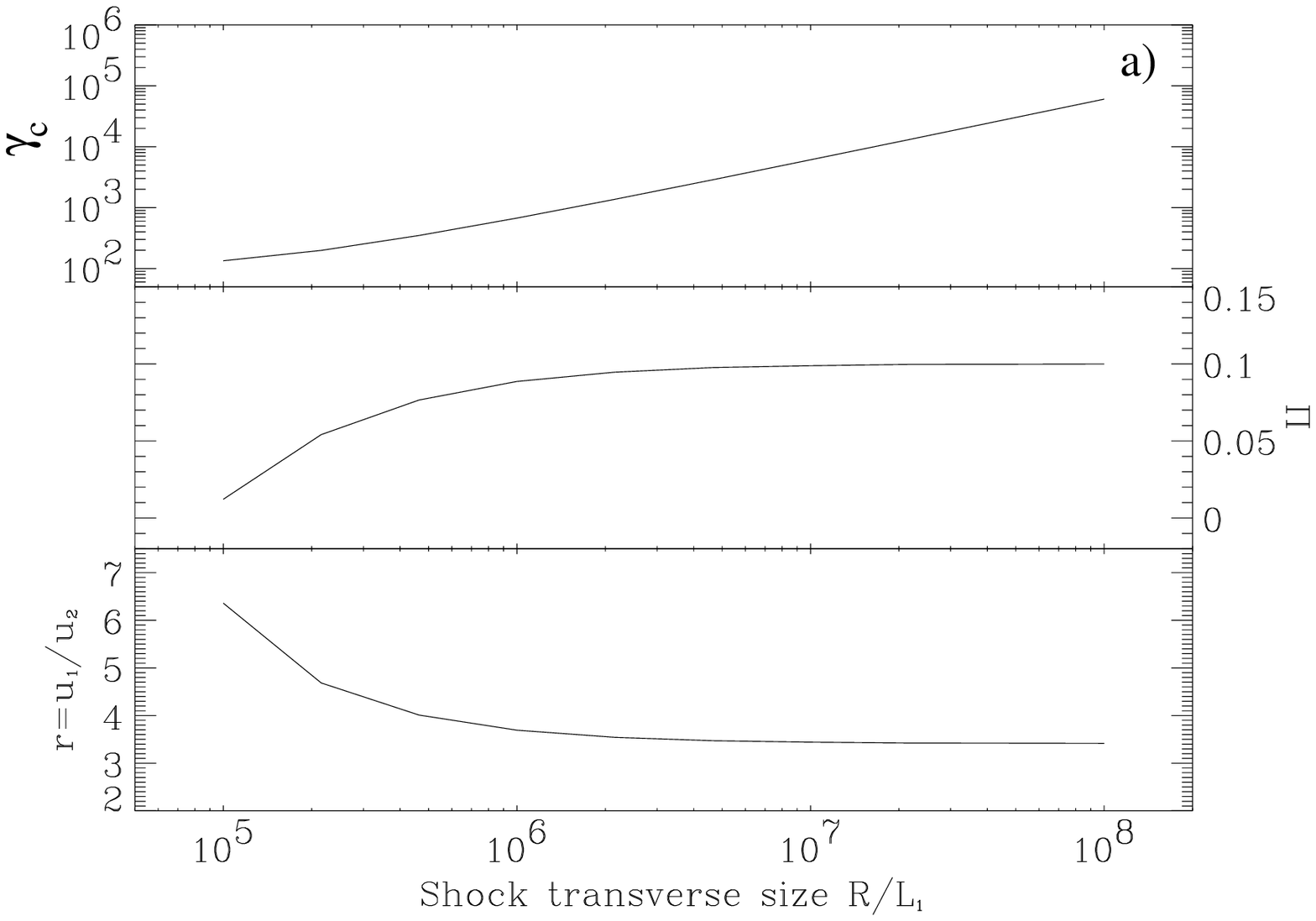}&\includegraphics[width=0.95\columnwidth]{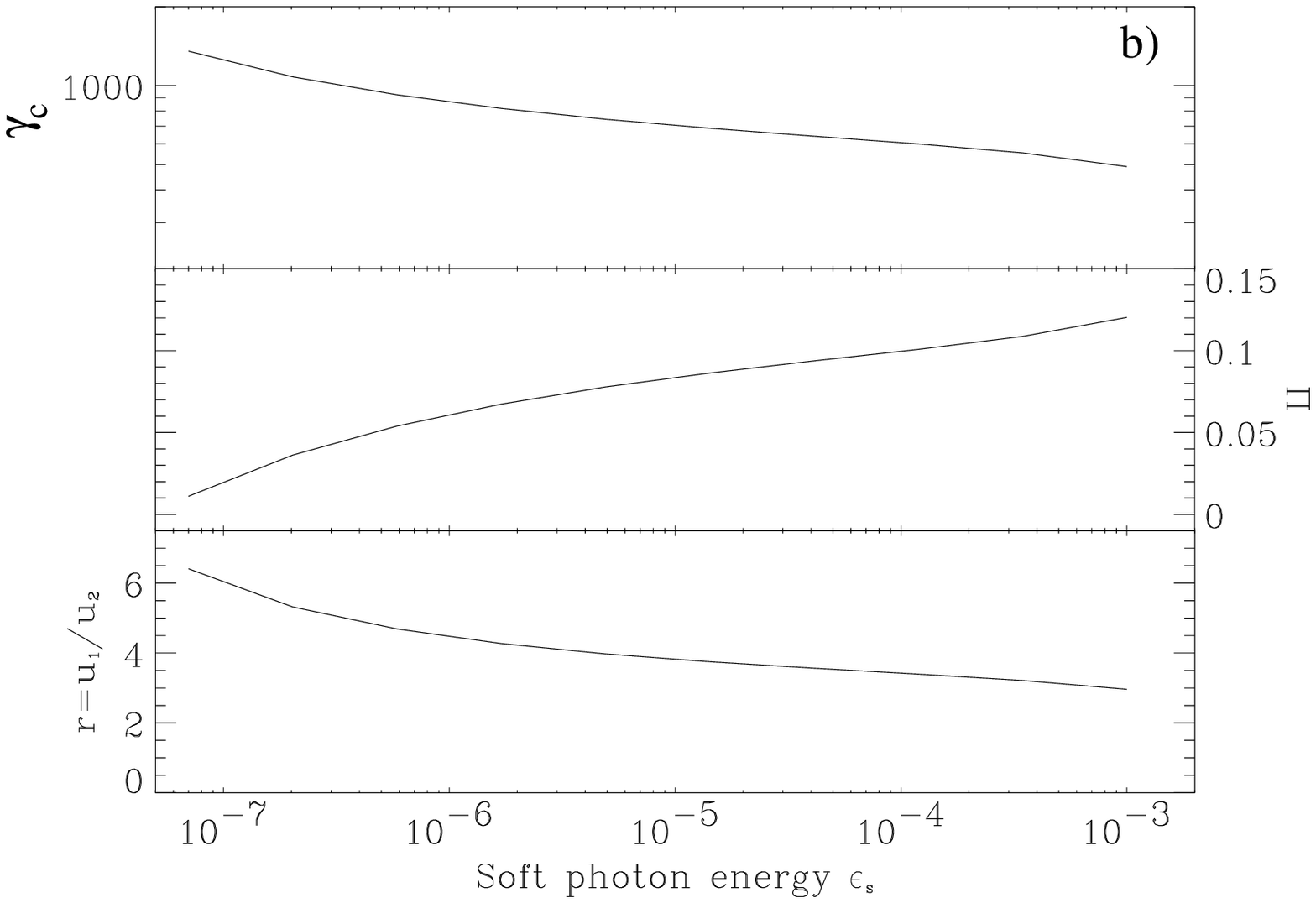}\\
&\\
\includegraphics[width=0.95\columnwidth]{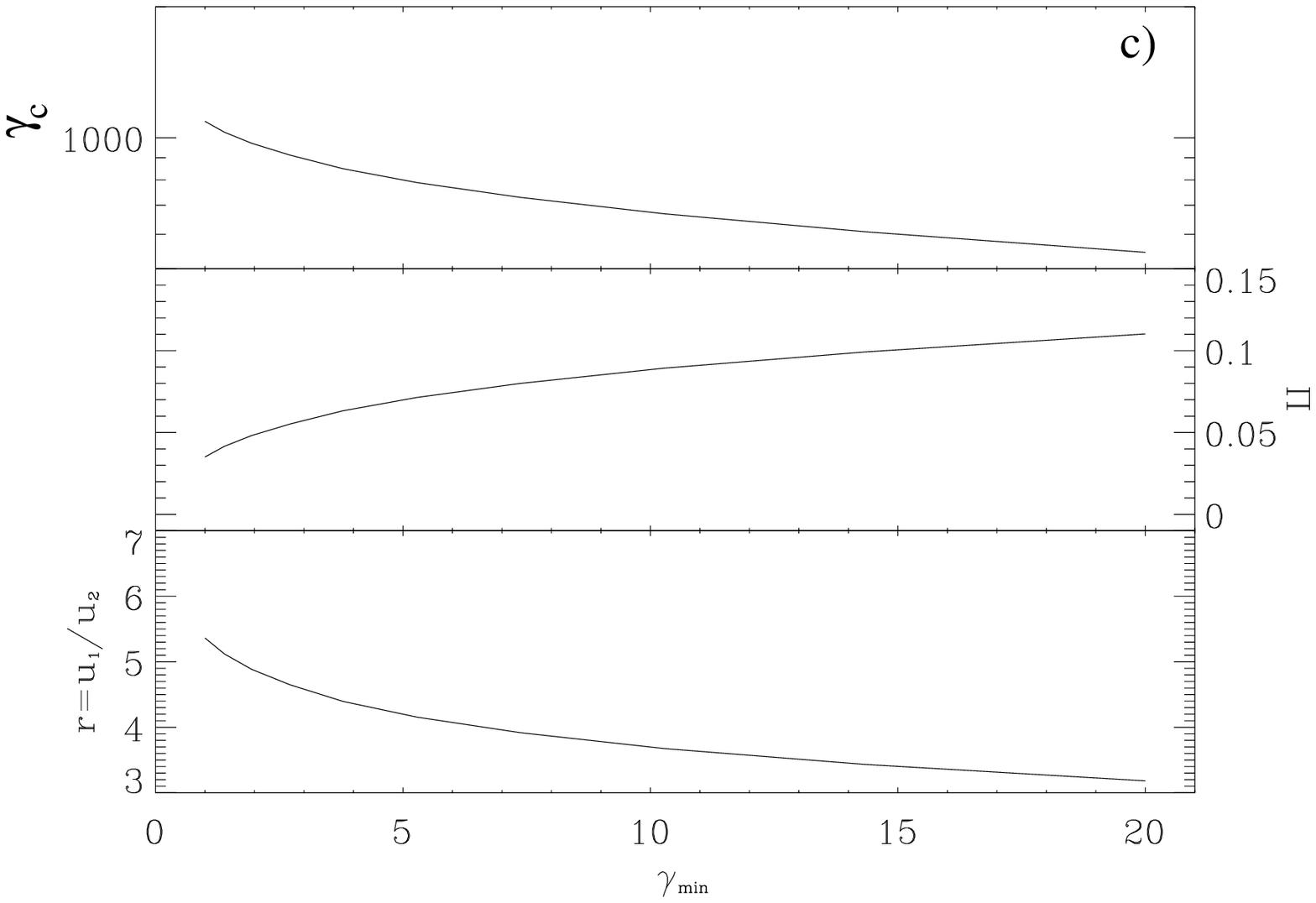}&\includegraphics[width=0.95\columnwidth]{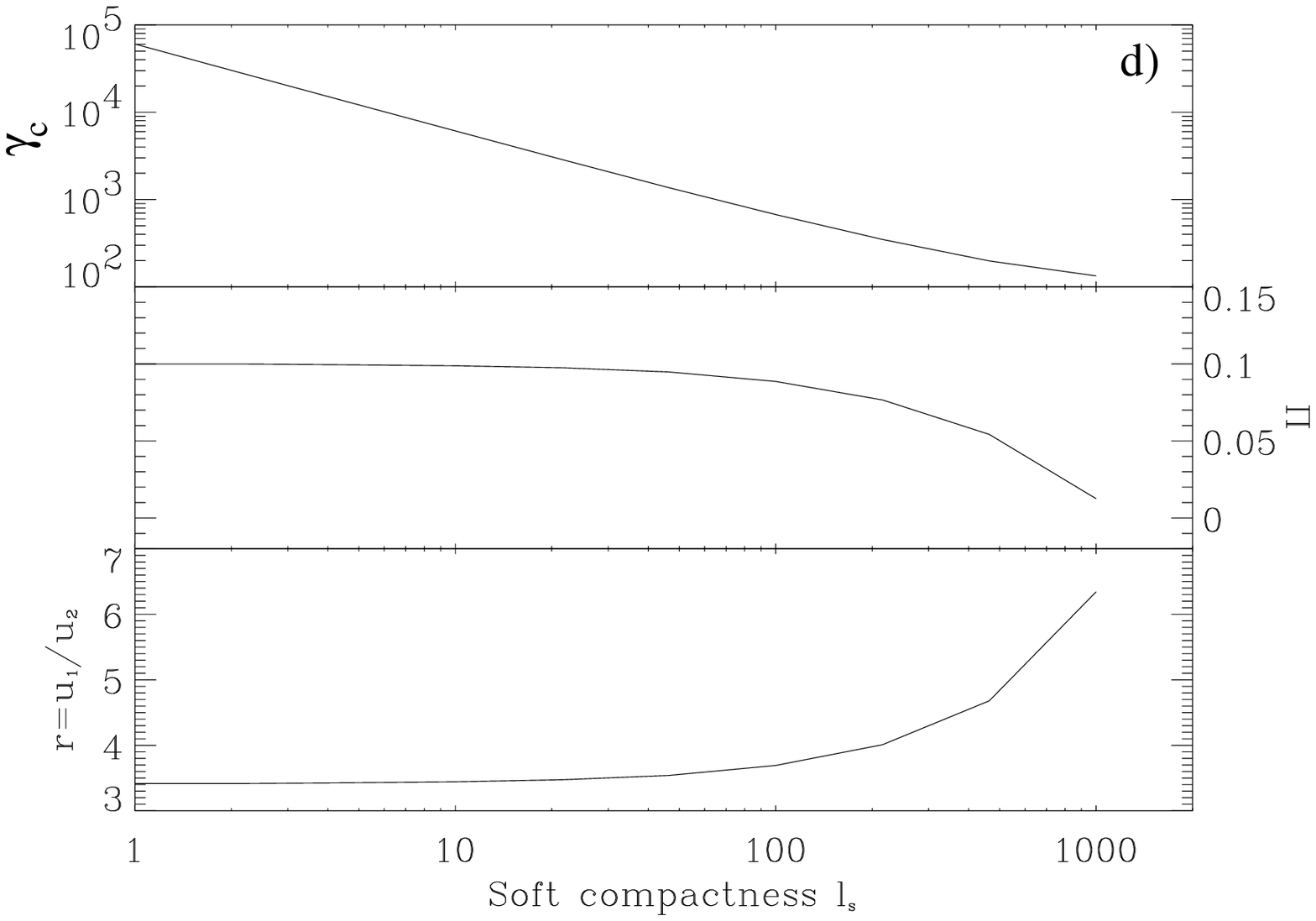}\\
&\\
\includegraphics[width=0.95\columnwidth]{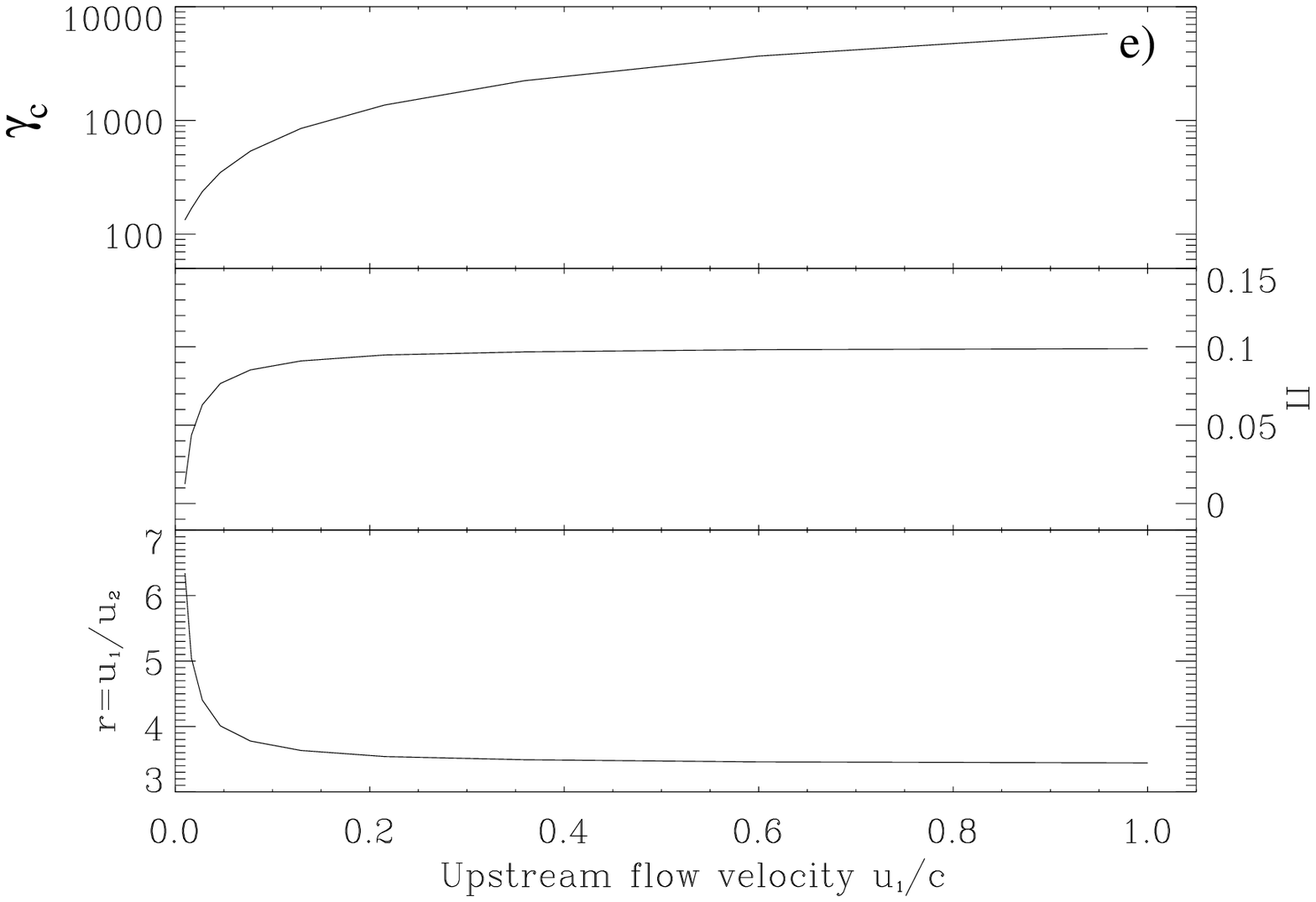}&\includegraphics[width=0.95\columnwidth]{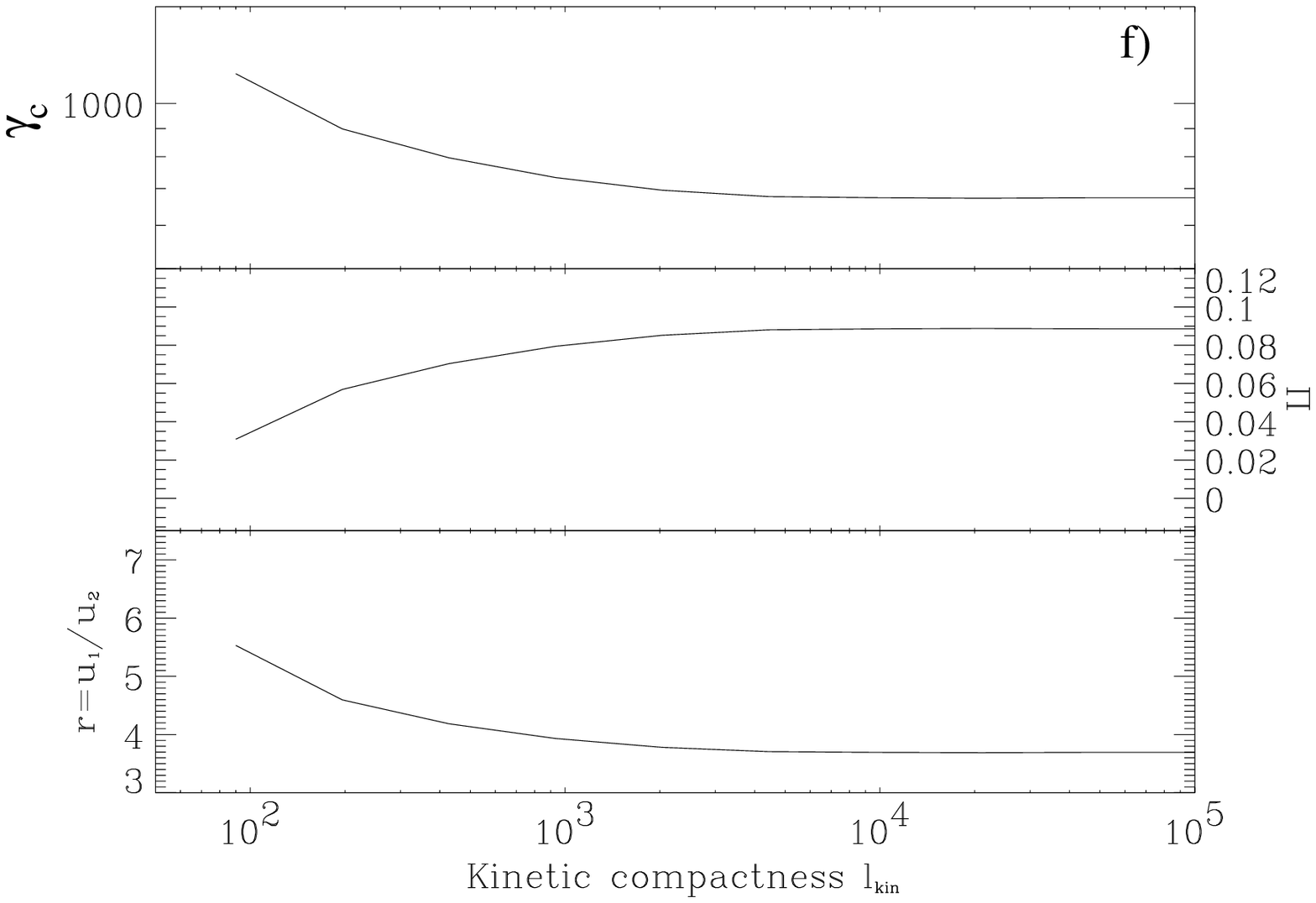}\\
\end{tabular}
\caption{Plots of $r$, $\Pi$ and $\gamma_c$ versus each parameter of the
model: a) the reduced shock transverse size $\wt{R}$, b) the reduced soft
photon energy $\epsilon_s$, c) the minimum Lorentz factor $\gamma_{min}$,
d) the soft compactness $l_s$ e) the upstream flow velocity $u_1/c$ and
f) the flow kinetic compactness the other parameters being fixed to
$\wt{R}=10^6$, $\epsilon_s=2\times 10^{-5}$, $\gamma_{min}=10$,
$u_1/c=0.1$, $l_s=100$ and $l_{kin}=10^4$.}
\label{figparamvar}
\end{figure*}
\end{center}

For clarity, and in order to better understand the influence of each
parameter on the solutions of our problem, we have plotted in Fig.
\ref{figparamvar}, the compression ratio $r$, the cut--off Lorentz factor
$\gamma_c$ and the pair luminosity $\Pi$ versus each one of these
parameters, the other being fixed to values indicated in the figure
caption. Some comments may be done on these different figures:
\begin{itemize}
\item Variation of $\wt{R}$: an increase of $\wt{R}$ will produce an
increase of $\wtRgg$ (cf. Eq. \ref{rggeq}) and thus an increase of $\Pi$
and a decrease of $r$. For $\wt{R}\gg L_o$, $\wtRgg$ converges to $L_o$
so that $\Pi$ and $r$ become constant (cf Fig.  \ref{figparamvar}a).  The
cut--off Lorentz factor $\gamma_c$ increases with $\wt{R}$ as expected
from Eq. (\ref{GAM}), because a constant $l_s$ implies a decreasing
photon density. For low values of $\wt{R}$, $\gamma_c$ become smaller
than $\gamma_{th}$ and the decrease of the pair creation rate must be
compensated by a hardening of the distribution ($r$ increases and $\Pi$
decreases).
  
\item Variation of $\epsilon_s$: when $\epsilon_s$ increases,
$\gamma_{th}$ decreases, favoring the pair creation process ($\Pi$
increases). It is compensated by a steepening of the particle
distribution (i.e. $r$ decreases). Concerning $\gamma_c$, it is a simple
function of $r$ (cf. Eq. \ref{GAM}), which decreases if $r$ decreases (cf
Fig.  \ref{figparamvar}b).

  
\item Variation of $\gamma_{min}$: since we suppose that the pairs are
  instantaneously accelerated to the Lorentz factor $\gamma_{min}$, the
  larger $\gamma_{min}$ and the larger the pair pressure i.e. the smaller
  the compression ratio (cf Fig.  \ref{figparamvar}c). We recall however
  that for too small values of $\gamma_{min}$ (i.e. $\gamma_{min}\simeq
  1$) the annihilation of pairs is not negligible anymore and our model
  is no more valid.
  
\item Variation of $l_s$: the larger the soft compactness, the shorter
  the cooling time scale (cf. Eq. (\ref{TOAb})) and thus the smaller
  $\gamma_c$. An increase of $l_s$ thus disfavors the pair creation
  process. The compression ratio has then to increase to keep the
  pair--hydrodynamics balance. For small $l_s$ however, $\gamma_c$ is
  very large ($\gg \gamma_{th}$) and its value becomes immaterial.
  Consequently, $\Pi$ and $r$, become independent of $l_s$, as shown in
  Fig. \ref{figparamvar}d.
  
\item Variation of $u_1$: when $u_1$ increases, $t_{acc}$ decreases and
then $\gamma_c$ increases. The compression ratio $r$ decreases by
compensation. On the other hand, for low values of $u_1$
(i.e. $u_1<0.1c$), $\gamma_c$ becomes too small, decreasing the pair
creation rate and $r$ increases rapidly.

\item Variation of $l_{kin}$: for high values of $l_{kin}$, the pair
  density must be also high to efficiently modify the hydrodynamical
  profile. The pair creation process is thus saturated meaning that the
  pair creation rate grows linearly with $l_{kin}$, i.e. $\Pi$, $r$ and
  $\gamma_c$ keep constant. However, for low values of $l_{kin}$, the
  pair density is so low that the pair creation optical depth becomes
  smaller than 1. In compensation $r$ must increase, so $\Pi$ decreases
  accordingly.
\end{itemize}


\section{Discussion}
In this paper, we have seen that the launch of the pair creation process
by particles accelerated by a shock embedded in a dense soft photon
field, could disrupt, through the increase of the associated pair
pressure, the plasma flow and eventually, for too high pressure, to
smooth completely the shock. By including the feedbacks of the
hydrodynamic of the flow on the pair creation process, the system can
reach, in some conditions, stationary states. Such processes may be at
the origin of the high energy emission observed in compact objects where
acceleration processes in dense photon field are expected to occur. We
give here numerical estimates of the physical parameters relevant to this
class of objects.

\subsection{High energy spectra}
\begin{figure*}[t]
  \begin{tabular}{ll}
\includegraphics[width=0.95\columnwidth]{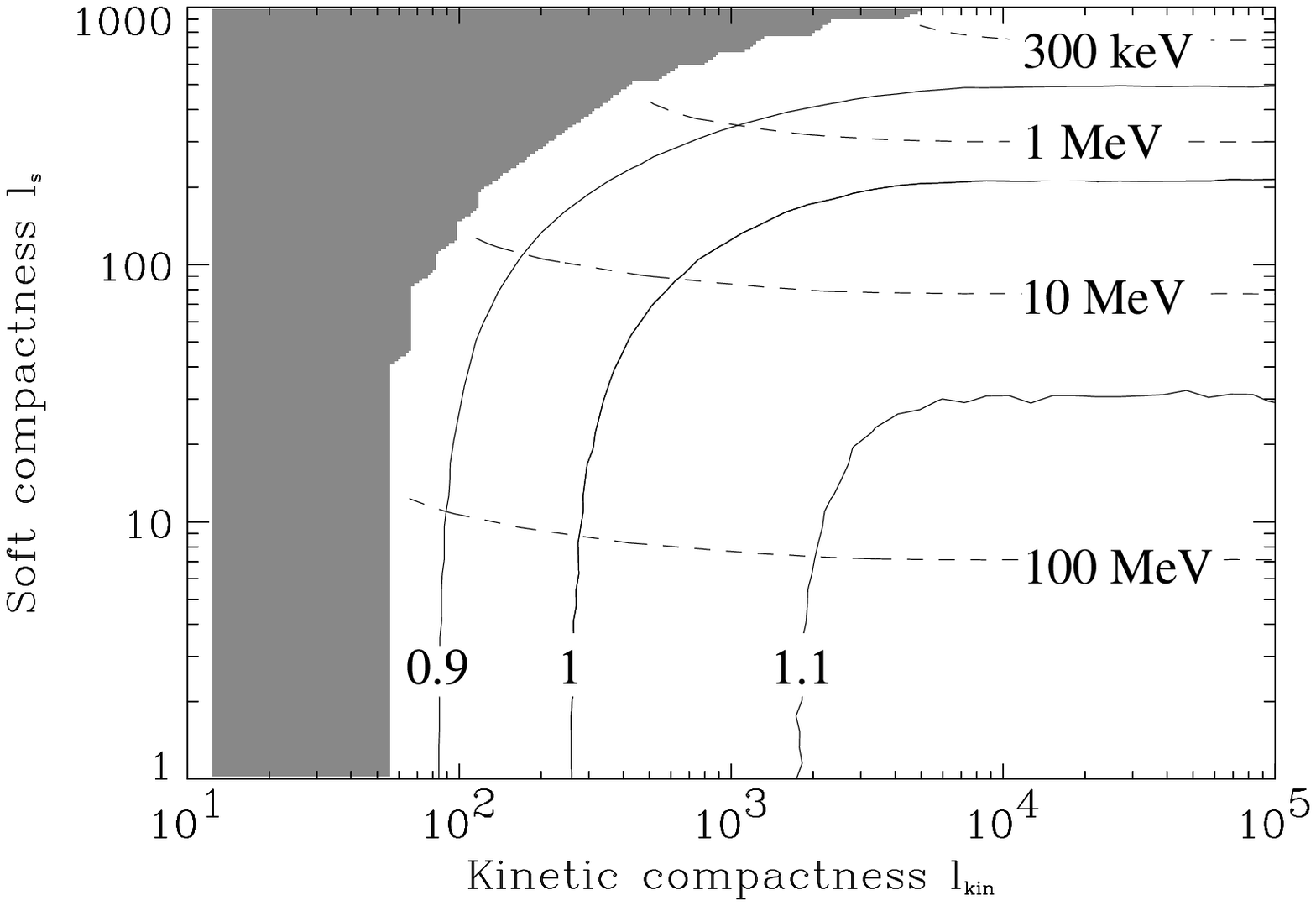}& 
\includegraphics[width=0.95\columnwidth]{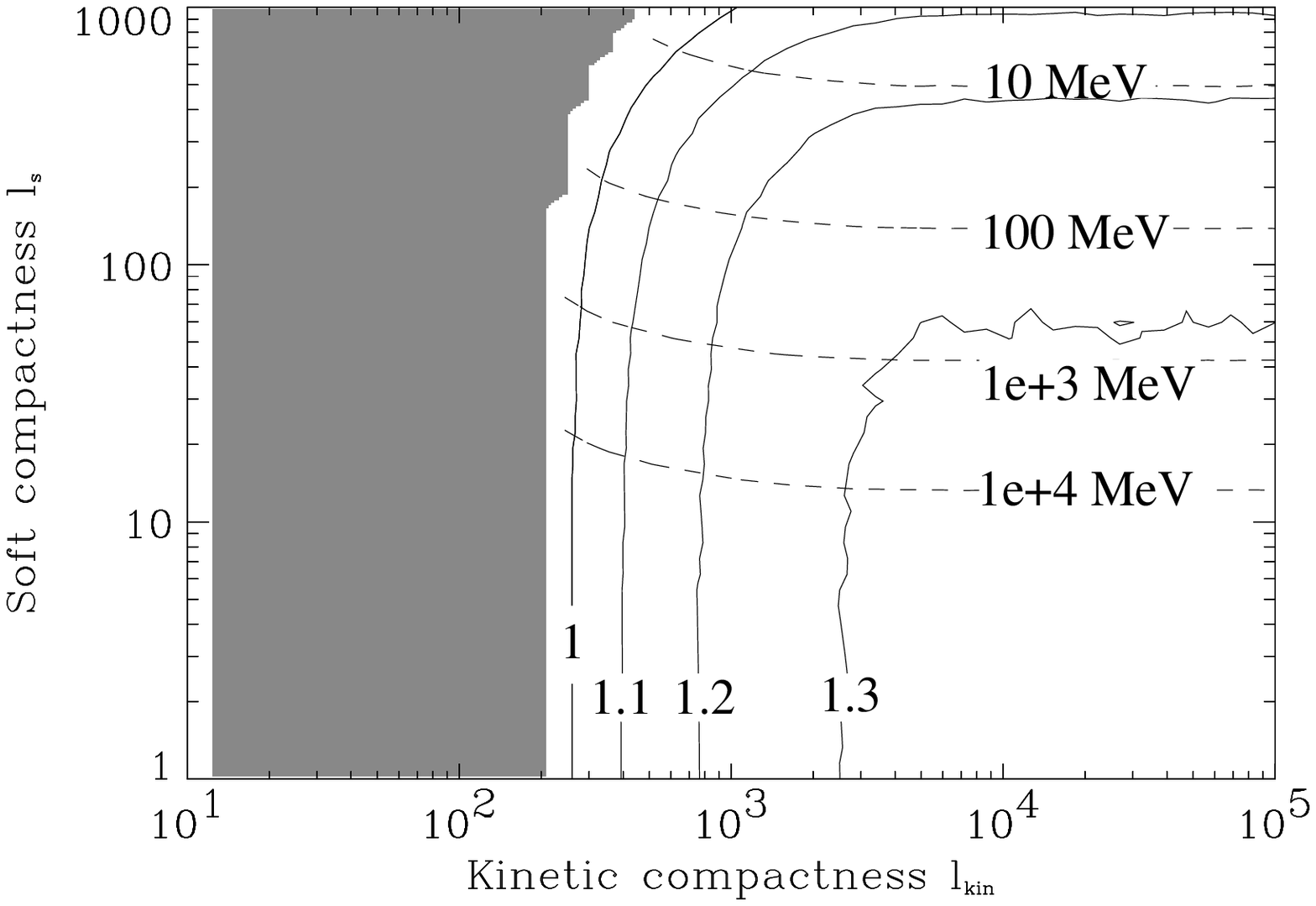}\\
\includegraphics[width=0.95\columnwidth]{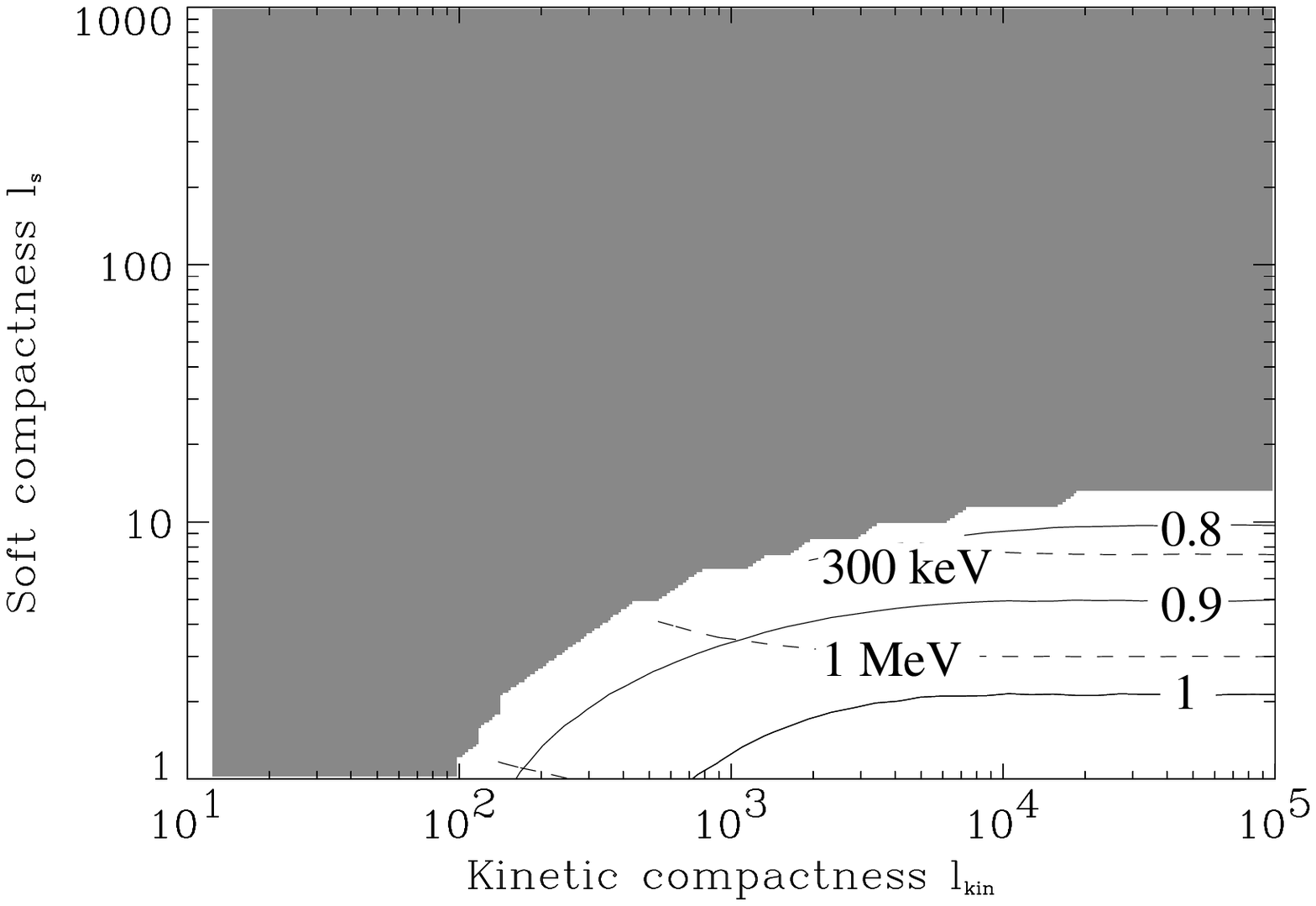}& 
\includegraphics[width=0.95\columnwidth]{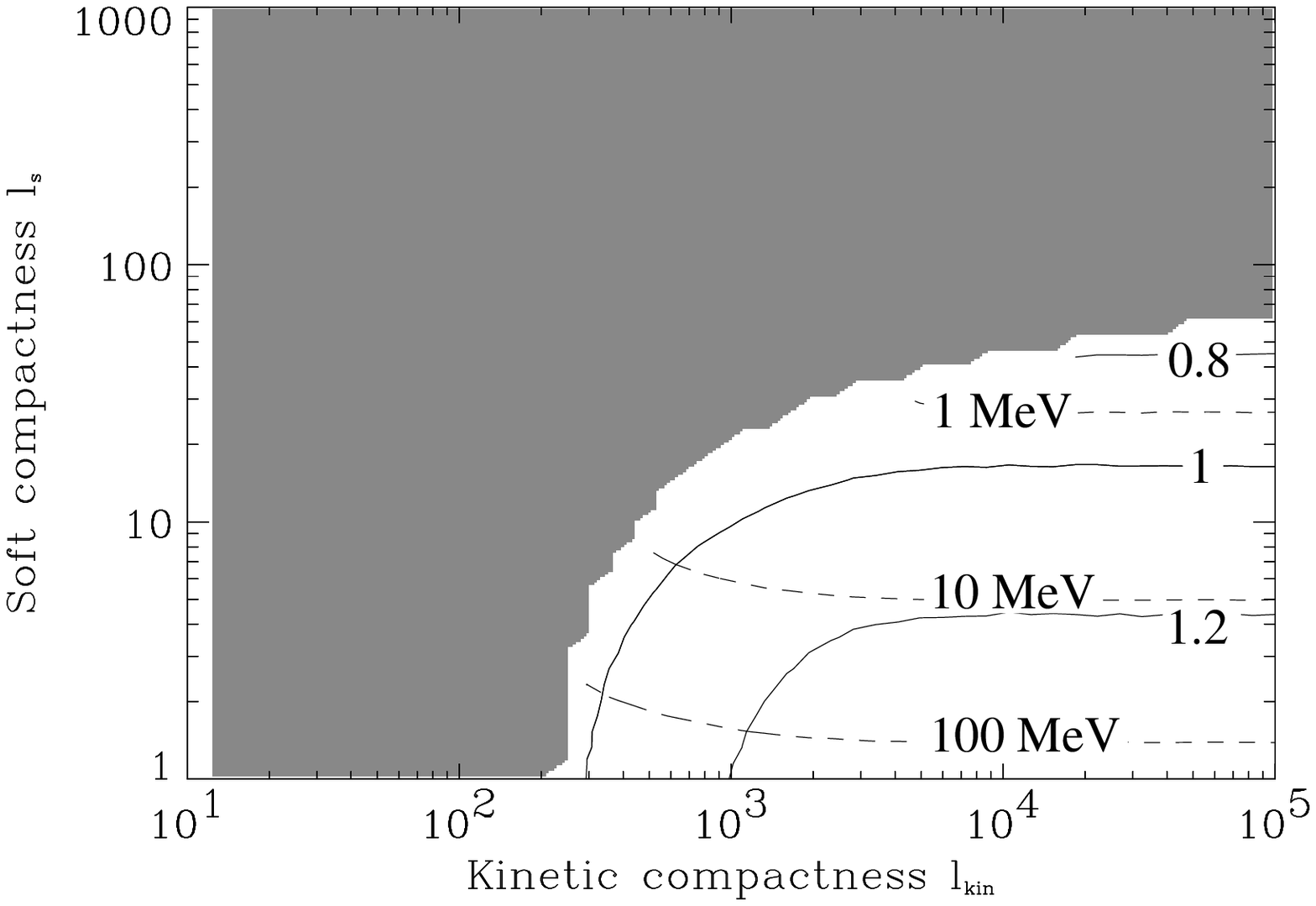}\\
  \end{tabular}
\caption{Contour plots of the spectral index $\alpha$ (solid lines) and
the high energy cut-off $E_c$ in keV (dashed lines) of the emitted
spectrum in the ($l_s$, $l_{kin}$) space. The soft photon energy is equal
to 10 eV and 600 eV for the left en right plots respectively and
$\wt{R}=10^4$ and $10^6$ for the lower and upper plots respectively. The
other parameters have been fixed to $\gamma_{min}=10$ and $u_1/c=0.1$. In
the grey region there is no pair dominated solution in stationary state .
\label{contours}}
\end{figure*}
As seen in the previous section, for a given set of parameters, the
system may reach stationary states characterized by a compression ratio
$r$. The spectral index and cut--off of the particle distribution
function just after the shock $n_2(\gamma,0^+)$ are then given by Eq.
(\ref{eqs}) and (\ref{GAM}) respectively. Since we have assumed that
these particles are cooled via IC process (where the cooling length scale
depends on the particle Lorentz factor), the emitted energy spectrum is
characterized by a cut-off power law shape $\displaystyle F_{E}\propto
E^{-\alpha}\exp \left[-\left(\frac{E}{E_c}\right)^{\frac{1}{2}}\right]$
(cf. Eq.  (\ref{eqspecph})) where $\alpha$ and $E_c$ are simple functions
of $s$ and $\gamma_c$:
\begin{eqnarray}
\alpha&=&\frac{s}{2}\\
E_c&\simeq&\frac{4}{3}\gamma_c^2\epsilon_sm_ec^2\label{eqEc}
\end{eqnarray}

We have reported in Fig \ref{contours} the contour plot of
$\alpha(l_s,l_{kin})$ and $E_c(l_s,l_{kin})$ for two values of
$\epsilon_s$ (10 and 600 eV which are representative of the typical
values of soft photons emitted by an accretion disk around a supermassive
and stellar mass black hole respectively) and $\widetilde{R}$ ($10^4$ and
$10^6$). The other parameters have been fixed to $\gamma_{min}=10$ and
$u_1/c= 0.1$. The contours of $\alpha$ and $E_c$ keep roughly the same
shape but cover a different region of the parameter space for different
parameter sets. In each figure, we can see that:
\begin{itemize}
\item The spectral index does not strongly vary between the harder and
the softer spectra ($\Delta\alpha\simeq0.3$). It reaches an asymptotic
plateau for low $l_s$ and high $l_{kin}$. In these conditions, both pair
density and high energy cut--off are large so that the pair creation
process is saturated. Then it can be shown that Eq. (\ref{eqX}) reduces
to a relatively simple relation between $\alpha$, $\gamma_{min}$ and
$\epsilon_s$, namely:
\begin{displaymath}
 \gamma_{min}^{1-2\alpha}=\frac{1}{2\alpha}\left(\frac{3}{4\epsilon_s}\right)^{1-\alpha}
\end{displaymath}
  Concerning $E_c$, following Eqs. (\ref{GAM}) and (\ref{eqEc}), it is
 inversely proportional to $l_s^2$.
\item The harder spectra are obtained for large values of $l_s$ or small
  values of $l_{kin}$. In both case, the pair efficiency decreases either
  due to a small value of $E_c$ or a low particle density, that is a
  low pair production optical depth. To keep the equilibrium between the
  pair creation effects and the hydrodynamic of the flow, the system has
  to reach harder spectra to compensate this decrease of the pair creation
  rate.
\item For small $l_s$ (respectively large $l_{kin}$), $\alpha$ is
independent of $l_s$ (respectively $l_{kin}$). This is a direct
consequence of the behavior of the compression ratio at small $l_s$ or
large $l_{kin}$ as explained in section \ref{solutions}. { We can note
that in the case $\epsilon_s =$ 600 eV and $\widetilde{R}=10^6$ shown in
Fig. \ref{contours} the condition $\gamma_c<m_ec^2/\epsilon_s$, needed in
Thomson regime (cf. section \ref{threg}), may not be verified in the low
part of the parameter space. However, since the pair creation process is
mainly triggered by particles with Lorentz factor of the order of
$\gamma_{th}$ ($\ll\gamma_c$), we do not expect the Klein-Nishina effects
to strongly modify our results.}
\item For to large $l_s$ or to small $l_{kin}$, the system cannot reach
sufficient hard states to keep in equilibrium and no high pair density
stationary state can exist any more. The system can only be in the
trivial pair free state (i.e. $r=7$ and $\Pi=0$).
\end{itemize}
We interpret the differences between the four plots of
Fig. \ref{contours} as follows. The increase of $\epsilon_s$ favors the
pair creation process. Thus, for given values of $l_s$ and $l_{kin}$, the
spectral index is larger. The high energy cut--off $E_c$ increases mainly
because of its dependence on $\epsilon_s$ (cf. Eq. \ref{eqEc}),
$\gamma_c$ keeping roughly constant (cf. Eq. \ref{GAM}). Concerning the
variation of $\widetilde{R}$, as expected, a larger value corresponds to
a more efficient acceleration (cf. Eq. \ref{GAM}) and favors the pair
creation process which allows a larger space parameter region for high
pair density solutions.

\subsection{Annihilation line}
The presence of pairs should give a signature as an annihilation feature
at $\sim$ 511 keV. We will show that this feature is not expected to be
strong in our model. Here, we have supposed the existence in the shock
region of pre--accelerating processes bringing leptons to the sufficient
energy (i.e. $\gamma > \gamma_{min}$ with $\gamma_{min}$ of the order of
a few, cf section \ref{accth}) for resonant scattering off magnetic
disturbances. Since the annihilation process occurs mainly at low energy,
i.e. for particles with Lorentz factor $\gamma\simeq 1$ (cf. Coppi \&
Blandford \cite{cop90}), it occurs mainly far downstream, where the pairs
created in the shock can cool down. The annihilation line luminosity is
thus at most equal to the pair rest mass luminosity, which is itself
smaller by a factor $\gamma_{min}$ than the pair creation luminosity (Eq.
\ref{pcrate}).  As shown in section \ref{maxpair}, the pair luminosity
$\Pi$ is itself limited and is necessarily smaller than $\sim$20\% of the
X-ray/$\gamma$-ray luminosity, i.e. $\Pi< \Pi_{max}\simeq 0.2$ (assuming
that the total kinetic energy of the upstream flow is transformed in
radiation). Besides, for $\Pi\simeq 0.2$, the compression factor is very
small, of the order of unity, resulting in a very steep X-ray spectra.
When hydrodynamics feedback is taken into account, the pair luminosity
may be well below this theoretical limit of 20\%. An X-ray spectral
photon index $\sim$ 2 (as those generally seen in Seyfert galaxies)
requires a compression ratio $r\sim$ 3--4. Such values of $r$ require
values of $\Pi$ smaller than $\simeq$ 10\% (cf. Fig \ref{rfuna}).\\

Assuming a steady state, pairs annihilate at the same rate as they are
produced. $\Pi/\gamma_{min}$ gives then an upper limit of the
annihilation line luminosity. We thus expect the luminosity of the
annihilation radiation to be smaller than few percent of the total high
energy radiation, which is quite compatible with the non observation of
strong annihilation lines in this class of Seyfert galaxies as shown by
the best upper limit observed in Seyfert galaxies with the
OSSE satellite (Johnson et al. \cite{joh97}).\\

\subsection{Variability}
For some values of external parameters, as suggested by Fig.
\ref{contours}, no high pair density solution can exist in stationary
states. So only the pair free solution (i.e. with $\Pi=0$) exists. The
system is not expected to be variable with a constant set of parameters
and variability can only occur with a variation of one of them. An
interesting possibility would be to consider a possible feedback of the
relativistic plasma to the soft compactness. In the reillumination models
for instance (Collin, 1991; Henri \& Petrucci, 1997), the soft photons
are produced by the reprocessing of the primary X-ray emission. An
increase of the pair plasma density will increase the X-ray illumination
and thus the soft compactness. Fig. \ref{contours} shows that in some
cases, the change of $l_s$ make the system switch to pair free solution
which will stop the reillumination and bring the system back to pair rich
solutions.  Limiting cycles could thus occur. We intend to further
investigate this possible effect under astrophysically relevant
conditions.

\section{Conclusion}
In the present paper, we have studied the effect of pair creation, via
high energy photon-photon interaction, on a shock structure, where the
high energy photons are produced via IC by the particles accelerated by
the shock itself. The problem is highly nonlinear since pairs can modify
the shock profile through their pressure and, mutually, a change of
the shock hydrodynamics can decrease or increase the pair production
rate.\\

We have shown that for a given size of the pair creation region,
it exist a maximal value of the pair creation rate above
which the shock cannot exist anymore.
When the hydrodynamical feedbacks on the pair creation process are
neglected, a pair power of at most 20\% of the upstream kinetic power is
sufficient to kill the shock. This constraint can fall to few percents in
stationary states where pair creation and hydrodynamical effects balance.
We thus do not expect the presence of strong annihilation lines. We also
obtain spectral parameters in rough agreement with the observations.\\

We suggest also a possible variability mechanism if the soft photon
compactness depends itself on the pair density of the hot plasma, such as
expected in reillumination models.\\

In the model presented here, the cooling of particles is due to the IC
process on external soft photons. However, particles may also cool on
soft photon they produce by synchrotron process when spiraling around the
magnetic field lines (the so-called synchrotron-self-compton process,
SSC). In this case the cooling will also depend on the particle
distribution function. We may expect that the addition of the SSC process
would allow to obtain stationary states with harder spectra than those we
obtained here, since the additional synchrotron cooling would be
compensate by a stronger acceleration, i.e. a larger compression ratio
$r$. The detailed study of this problem is left to future work.

\begin{acknowledgements}
We acknowledge the anonymous referee for his helpful comments. POP
acknowledges a grant of the European Commission under contract number
ERBFMRX-CT98-0195 (TMR network "Accretion onto black holes, compact stars
and protostars").
\end{acknowledgements}

\appendix
\section{Particle distribution function with pair creation}
\label{app1}
Neglecting the annihilation rate which is only important at low energy,
and asssuming that pairs are produced at a constant rate $Q$ in the
region $-\Rgg\leq x\leq +\Rgg$ and instantaneously accelerated at
$p_{min}=\gamma_{min}m_ec^2$ (cf. section \ref{accth}), Eq. (\ref{eqfb})
can be rewritten (in the stationnary state) as follows:
\begin{eqnarray}
  \dt{f}+u\dx{f}&=&\frac{1}{3}\dx{u} p
  \dpp{f}+\frac{1}{p^2}\dpp{bp^4f}+\dx{}D\dx{f}+\nonumber\\
   & &  Q\delta(p-p_{min})H (\Rgg-|x|)
  \label{eqfbb}
\end{eqnarray}
where $H(x)$ the Heaviside function defined by
\begin{eqnarray*}
H (x)&=&1 \mbox{ for } x\ge 0\\
&=&0 \mbox{ everywhere else. }
\end{eqnarray*}
As shown in section \ref{pairreg}, the shock thickness (which is of the
order of the diffusion length) is largely smaller than any of the
characteristic length of the problem. Consequently, when integrating Eq.
(\ref{eqfbb}) between $-\infty$ and $0^+$, we can still approximate the
velocity gradient by a Dirac $\displaystyle\dx{u}=(u_2-u_1)\delta (x)$.
As we can also neglect the coolings for $x\leq 0$ for particles with
Lorentz factor $\gamma\le\gamma_c$ (cf. section \ref{accreg}). The result
of the integration gives:
\begin{eqnarray}
f_{0^+}+\left (\frac{r-1}{3r}\right
)p\dpp{f_{0^+}}&=&f_1+\frac{Q\Rgg}{u_1}\delta
(p-p_{min}) \nonumber\\
 & & +\frac{D}{u_1}\dx{f_{0^+}}
\label{eqapp}
\end{eqnarray}
where $f_{0^+}$ and $f_1$ are the distribution function in $0^+$ and in
$-\infty$ respectively. In the downstream flow only the pair creation
process will modify the distribution particles, and the main result will
be an increase of the total number of particles on a scale of the order
of $\Rgg$. We can thus make the following approximation:
\begin{displaymath}
\frac{D}{u_1}\dx{f_{0^+}}\simeq
\frac{D}{u_2}\frac{u_2}{u_1}\frac{f_{0^+}}{\Rgg}\simeq
\frac{L_2}{\Rgg}\frac{f_{0^+}}{r}
\label{eqeq}
\end{displaymath}
Since we have $\Rgg\gg L_2$, this term is 
negligible in Eq.~(\ref{eqapp}),
and its integration gives:
\begin{displaymath}
f_{0^+}(p)=qp^{-q}\int_{0}^{p}\left[f_1(p')+\frac{Q\Rgg}{u_1}\delta(p'-p_{min})\right]{p'}^{q-1}dp'
\end{displaymath}
With $Q=0$, one finds again the solution without pair creation. For
$p>p_{min}=\gamma_{min}m_ec$, the integral is almost constant. We thus
find that the distribution function is a power law of spectral index
$\displaystyle q=\frac{3r}{r-1}$ in any case.

%

\end{document}